\documentclass[a4paper,11pt]{article}

\usepackage{aaskaiid}
\usepackage{verbatim}
\usepackage{enumerate}
\usepackage{orcidlink}    

\title{The Role of Artificial Intelligence in the SKA Era}
\ShortTitle{AI in the SKA era}

\author[1]{Philipp Denzel~\orcidlink{0000-0003-0126-0659}}
\ShortName{Denzel et al.} 
\author[1]{Frank-Peter Schilling~\orcidlink{0000-0002-7763-2140}}
\author[1,2]{Elena Gavagnin~\orcidlink{0000-0001-5146-3759}}

\affiliation[1]{Centre for Artificial Intelligence, Zurich University of Applied Sciences ZHAW, Technikumstrasse 71, Winterthur 8400, Switzerland}
\emailAdd{denp@zhaw.ch}
\emailAdd{scik@zhaw.ch}
\affiliation[2]{Institute of Business Information Technology, Zurich University of Applied Sciences ZHAW, Theaterstrasse 17, Winterthur 8400, Switzerland}
\emailAdd{gava@zhaw.ch}

\abstract{
  The Square Kilometre Array Observatory (SKAO) will usher in an era of unprecedented data complexity and scientific opportunity in radio astronomy, producing petabyte-scale datasets and terabit-per-second streams that challenge traditional analysis paradigms.
  Artificial Intelligence (AI) stands at the forefront of this transformation, offering scalable, adaptive solutions to the most pressing problems in radio astronomy and astrophysics. This chapter explores the pivotal role of AI in the SKA era, from real-time operations to scientific discovery.
  We examine how deep learning models enable automated source detection, radio-frequency interference mitigation, anomaly detection, and parameter inference, while generative approaches accelerate sky simulations, calibration, and imaging.
  Reinforcement learning promises dynamic scheduling and autonomous system control, and federated learning could address the distributed nature of SKA data.
  Beyond performance, we emphasize the necessity of explainability, uncertainty quantification, and physics-informed inductive biases to ensure scientific integrity.
  By mapping SKAO's core challenges -- data volume, complexity, and interpretability -- onto modern AI methodologies, we review how deep learning, self-supervised frameworks, and probabilistic models can unlock new frontiers in cosmology, galaxy evolution, and time-domain astrophysics.
  AI is not merely an automation tool for coping with scale. It is a catalyst for discovery, redefining how we observe, model, and understand the Universe.
}

\begin{document}
\maketitle

\section{Introduction}

The Square Kilometre Array Observatory (SKAO) will transform radio astronomy by delivering orders-of-magnitude gains in sensitivity, angular resolution, and survey speed across a wide range of radio frequencies.
With that leap comes a scale and complexity shock: terabit-per-second raw data rates streaming off the arrays, exabyte-scale archives, and a heterogeneous landscape of observing modes (calibrated visibilities, interferometric imaging, beam-formed products) that stress every step of the scientific workflow -- from calibration and imaging to inference and discovery.
Traditional pipelines meticulously engineered around manual parameter tuning, human-in-the-loop quality control, and post-facto archival mining are not designed to function in real time, nor to maintain statistical optimality in the face of rapidly varying instrumental and environmental conditions.
Artificial Intelligence (AI) -- encompassing supervised, self-/semi-supervised, generative, probabilistic, reinforcement, physics-informed, and federated paradigms -- has matured precisely as SKA-class challenges crystallize.
Across precursor and pathfinder experiments and data challenges, AI has already demonstrated the ability to denoise, detect, classify, reconstruct, emulate, and infer at scale while quantifying uncertainty and respecting physical constraints~\citep[e.g.,][]{bonaldi_2018_square_kilometre, bonaldi_2025_square_kilometre, lastufkaSelfSupervisedLearningMeerKAT2024, riggiAstronomicalSourceDetection2023, acharya21cmSignalEpoch2024}.
The rationale for a broad AI integration within SKA is therefore twofold:

\begin{enumerate}
  \item \textbf{Operational necessity}: automate (near) real-time decision-making and data triage, preserve faint signals under extreme compression, and stabilize calibration/imaging under direction-dependent and time-variable systematics.
  \item \textbf{Scientific leverage}: extract non-Gaussian information, accelerate forward modelling, and deliver uncertainty-aware inferences at the field level; enabling robust cosmology, galaxy evolution, and time-domain astrophysics.
\end{enumerate}

This chapter underscores AI as a central component to SKAO's operational viability and scientific reach.
We categorize the SKAO's problems, list the failure modes of status-quo approaches, and propose AI-driven remedies that emphasize explainability, uncertainty quantification, and physics-aware inductive biases.
We also outline Machine Learning Operations (MLOps) and governance practices required for reproducible, auditable data science at exascale.
Finally, we review AI techniques for forward sky modelling, ranging from generative models for cosmological light cones to physics-informed emulation of baryonic processes.
These methods are indispensable accelerators that enable rapid hypothesis testing and bridge the gap between theory and observation at SKA scales.

Building on this foundation, we now turn to a selection of specific scientific domains where AI offers transformative capabilities. The following section outlines key problem areas in SKA science, ranging from early-Universe cosmology to time-domain astrophysics, where traditional approaches face fundamental limitations, and where AI-driven methods promise significant advances in sensitivity, scalability, and interpretability.

\section{Scientific problem areas}\label{sec:scientific_problems}

The scientific opportunities enabled by the SKAO come with analysis regimes that stress or break traditional methods: extreme data volumes and rates, highly non-Gaussian signals, complex instrument-sky systematics, and the need for real-time decision-making.
In this Section, we identify problem classes in astrophysics and radio astronomy where AI, in particular, deep learning methods have the potential to materially outperform more traditional pipelines; either by improving sensitivity and robustness, reducing latency and cost, or enabling/automating analyses that are otherwise impractical.
For each area, we outline
\begin{enumerate}[(i)]
  \item the scientific objective and limitations of status-quo approaches,
  \item AI-based strategies likely to deliver leaps in performance, referencing a selection of recent works implementing such strategies,
  \item followed by risks and challenges induced through the application of AI, e.g., uncertainty quantification, interpretability, and physics-aware inductive biases.
\end{enumerate}
Technical definitions of the AI methods referenced throughout this section are collected in Appendix~\ref{sec:ai_methods} for reference.


\subsection{Radio interferometric imaging \& sky simulations}\label{sec:radio_interferometric_imaging}

Radio interferometric imaging is governed by the radio interferometer measurement equation (RIME), an inherently ill-posed inverse problem where minor perturbations in visibility data can produce substantial deviations in reconstructed images.
For decades, the CLEAN algorithm~\citep{wright_2024_clean_algorithm} has dominated this domain, relying on the heuristic assumption of a sparse sky composed primarily of point sources.
While widely adopted, CLEAN exhibits critical limitations for exascale data requirements of the SKA era: poor computational scalability with image size and source count, dependence on manual expert intervention for convergence, reduced accuracy for faint or diffuse structures, and the absence of statistically calibrated confidence measures.
More modern variants such as SARA \citep[Sparsity Averaging Reweighted Analysis;][]{carrillo_2012_sparsity_averaging} improve reconstruction fidelity but remain computationally prohibitive. Recent developments, such as Bluebild~\citep{tolley_2025_bipp}, leverage eigendecomposition and GPU acceleration to reduce time-to-solution for large-scale imaging, yet fundamental challenges persist~\citep[see also][]{LeZhang01.2026.SKA}.

To benchmark next-generation algorithms, realistic sky simulations have become indispensable.
Simulation frameworks like OSKAR~\citep{dulwich_2017_oskar} and Karabo~\citep{sharma_2026_karabo} which produce synthetic interferometric visibilities in the $uv$-plane by modelling a full signal chain, enable end-to-end simulation of telescope characteristics, including beam patterns, system noise, calibration errors, and ionospheric distortions.
Complementary hydrodynamical simulations -- most relevant for \textit{SKA-Mid} science at low to intermediate redshifts -- provide physically motivated mock skies incorporating diffuse emission, extended structures, and multi-scale interactions, far surpassing simplistic point-source or Gaussian models in complexity.
Magneto-hydrodynamics are essential for capturing magnetic-field-driven phenomena such as synchrotron radiation, Faraday rotation, and polarization.
However, a persistent computational tension exists between the need for cosmological volumes to sample rare sources and the sub-kpc resolution required to match \textit{SKA-Mid}'s sub-arcsecond imaging capabilities.
Current ``zoom-in'' simulations achieve the necessary resolution for phenomena like jet formation and turbulence but only within limited volumes.
Moreover, sensitivities in subgrid physics tuning parameters exacerbate the computational burden, underscoring the need for efficient emulation strategies.

\paragraph{Interferometric deconvolution} is increasingly utilizing AI-based solutions to these challenges, using deep learning architectures such as convolutional neural networks (CNNs; Appendix~\ref{sec:cnn}), U-Nets (Appendix~\ref{sec:unet}), and residual networks (e.g. ResNets; Appendix~\ref{sec:resnet}).
These models aim to accelerate processing, enhance reconstruction fidelity for extended emission, and mitigate artefacts in dirty images.
For instance, AIRI~\citep[AI for Regularization in radio-interferometric Imaging;][]{terris_2022_image_reconstruction} integrates a neural denoise within an optimization framework, achieving image quality comparable to SARA at significantly reduced computational cost.
\cite{schmidt_2022_deep_learning} explore end-to-end U-Net architectures for direct visibility-to-image reconstruction, trading extreme inference speed against risks of hallucinations and poor generalization.
POLISH~\citep{connor_2022_deep_radio}, a ResNet-based super-resolution model, designed to ``polish'' dirty images, improves signal-to-noise ratio (SNR) by 3 dB over CLEAN and is tailored for high-throughput surveys such as DSA-2000.
The R2D2 \citep[Residual-to-Residual DNN series of high-Dynamic range imaging;][]{aghabiglou_2024_r2d2_deep, aghabiglou_2025_toward_robust, tajja_2025_s_r2d2} paradigm replaces heuristic steps with iteration-specific neural networks, delivering high-dynamic-range imaging at CLEAN-like speeds.
Semi-supervised frameworks such as VisRec~\citep{wang_2025_visrec} further enhance robustness under limited labelled data.
Conditional generative diffusion models \citep[Appendix \ref{sec:diffusion}, e.g.][]{drozdova_2024_radio_astronomical} achieve superior source localization and flux estimation, while \cite{martinezSimulatingImagesRadio2024} demonstrate conditional diffusion models as emulators for realistic synthetic skies, enabling controlled morphology and flux generation for training and benchmarking.

\paragraph{Radio-frequency interference (RFI) mitigation and calibration} traditionally major obstacles to interferometric imaging, are typically treated as a detection or segmentation task by deep learning models.
RFI mitigation is critical for both \textit{SKA-Mid} and \textit{SKA-Low}, though the challenges differ: SKA-Mid contends primarily with terrestrial transmitters and satellite interference across 350\,MHz--15\,GHz, while SKA-Low operates in a crowded low-frequency band (50--350\,MHz) where RFI is pervasive and spectral purity is essential for EoR science.
For both, U-Nets and transformer-based architectures (Appendix~\ref{sec:transformer}) are primarily employed, but more customized hybrid approaches have been explored, combining neural networks with wavelet-based decomposition.
Notable examples include convolutional U-Nets for RFI detection~\citep{akeret_2017_radio_frequency}, residual architectures such as R-Net for MeerKAT \citep{sadr_2020_deep_learning}, and liquid state machines, a type of spiking neural network that processes temporal spike patterns rather than continuous activations~\citep{pritchard_2025_advancing_rfi}, achieving near-perfect per-pixel accuracy.
Recent work leverages foundation models like SAM~\citep{deal_2024_segmenting_rfi} for automated RFI segmentation without manual intervention.

\paragraph{Cosmological hydrodynamical simulations} proxy for theory, and can be used as tests against observations.
Since simulations track the evolution of gas density, velocities, and temperature, which telescopes do not observe directly, it is essential to have translations that enable a comparison of simulated and observational quantities~\citep[see also][]{Lagos01.2026.SKA}.
This is non-trivial and requires complex models such as conditional generative models for map-to-map translation (see Appendices~\ref{sec:cond_generation}, \ref{sec:cgan}, and \ref{sec:diffusion}).
\cite{denzel_2025_galactic_alchemy} propose such a solution, primarily relevant for \textit{SKA-Mid} science, which probes low-redshift HI and galaxy structure~\citep[see also][]{JingWang01.2026.SKA}.
The core premise is that the mappings between various simulated quantities of low-redshift galaxies such as gas to neutral hydrogen (HI) brightness or dark matter, while complex, are deterministic enough to be learned by a neural network.
The study comparatively investigates the computational efficiency and accuracy of two state-of-the-art paradigms GANs and diffusion models, using physics-aware metrics to assess observational plausibility.
By tuning the U-Net architecture, this work reduces computational cost, and bridges the gap between scalable, generative frameworks for forward modelling and observational reconstruction.
{EMBER} \citep{bernardini_2026_ember_hi, bernardini_2025_ember_2, bernardini_2021_from_ember} exemplifies this approach, accurately predicting key statistics such as mass fractions and profiles, and recovering HI-dark matter correlations across cosmic time down to smaller scales.
At higher redshifts, where full hydrodynamical simulations become computationally prohibitive over the cosmological volumes required by SKA-Low, semi-numerical codes such as 21cmFAST~\citep{mesinger_2010_21cmfast} or BEoRN~\citep{schaeffer_2023_beorn} provide the tractable simulation backbone for forward modelling.
Neural emulators trained on such codes -- most notably 21cmEMU~\citep{breitman_2023_21cm_emu} -- can reproduce the 21-cm power spectrum, global signal, mean neutral hydrogen fraction $x_\text{HI}(z)$, and UV luminosity functions across a multi-dimensional astrophysical parameter space with median errors below 1\% and a speed-up of roughly 10$^4$ over the underlying simulator.

Despite these advances, significant challenges remain.
Deep learning-based deconvolution is inherently opaque, raising concerns about hallucinated structures and bias propagation from training data.
The absence of calibrated uncertainties further complicates scientific validation.
Bayesian frameworks such as quantifAI~\citep{liaudat_2023_scalable_bayesian} address these issues by embedding learned priors within convex constraints, ensuring log-concave posteriors and enabling pixel-wise uncertainty estimation.
Physics-informed inductive biases and rigorous benchmarking on realistic simulations are essential to mitigate domain gaps and enhance interpretability.
Finally, robust MLOps practices (Appendix~\ref{sec:mlops}), including data versioning, audit trails, and continuous monitoring, are critical for reproducibility and long-term reliability of AI-driven imaging and simulation pipelines at SKA scale.


\subsection{Epoch of Reionization and 21-cm Cosmology}\label{sec:eor}
Detecting the redshifted 21-cm signal from the Epoch of Reionization (EoR), the Cosmic Dawn, represents one of the most challenging objectives in observational cosmology.
It is a main scientific target of \textit{SKA-Low}, which will observe the 21-cm line across the EoR in its AA4 and AA$\ast$ configurations~\citep[see also][]{Trott01.2026.SKA}.
The task involves isolating a faint, highly non-Gaussian cosmological field from spectrally smooth but instrumentally distorted foregrounds that are four to five orders of magnitude brighter, while accounting for contributions such as direction-dependent beams, calibration errors, and thermal noise.
Classical approaches, polynomial foreground fitting, principal component analysis, wedge avoidance in $(k_\perp, k_\parallel)$ space, and power-spectrum estimators, prioritize robustness at the expense of sensitivity.
These methods often suffer from leakage due to calibration residuals or, conversely, excessive foreground suppression that erases cosmological signal components, introducing systematic biases.
Furthermore, most traditional pipelines neglect field-level inference, forfeiting morphological information critical for constraining astrophysical processes.

\paragraph{Foreground separation} beyond linear models has emerged as a key application of AI~\citep[see also][]{Spinelli01.2026.SKA,Burba01.2026.SKA}.
Self-supervised and semi-supervised representation learning enables models such as variational auto-encoders (VAEs), diffusion models, or normalizing flows to learn the manifold of smooth foregrounds jointly with instrumental effects (see Appendices \ref{sec:vae}, \ref{sec:diffusion}, and \ref{sec:norm_flow}).
These artificial neural network (ANN) architectures can be embedded within Bayesian frameworks to provide uncertainty-aware inference.
Conditional generative models (Appendix~\ref{sec:cond_generation}) further allow projection of visibilities or image cubes onto foreground subspace while preserving cosmological modes, by modelling the conditional distribution $p(\textrm{foreground}\,|\,\textrm{observation})$ (treating cosmological signals as residuals).
\cite{acharya21cmSignalEpoch2024} for instance, introduce a VAE-trained kernel for Gaussian Process Regression (GPR) foreground mitigation, addressing signal loss due to misestimated covariance.
Tested on LOFAR-like simulations, this approach improves recovery of both shape and power of the 21-cm signal.
Subsequent work~\citep{acharyaRevisedLOFARUpper2024} applies this methodology to 141 hours of LOFAR data, reporting revised upper limits and reduced correlation with systematic excess noise; similar GPR approaches have been studied recently by \cite{ceccottiFirstUpperLimits2025b, mertensDeeperMultiredshiftUpper2025a, liuBayesianModelComparison2025}.
Deep learning-based foreground mitigation has also been explored using convolutional architectures.
\cite{gaoExtractingEpochReionization2025a} employ a 3D U-Net to extract EoR signals from SKA-Low simulations (AA$\ast$) contaminated by foreground residuals and thermal noise, achieving reliable power-spectrum predictions at long integration times.
Similarly, \cite{liSeparatingEoRSignal2019} demonstrate that convolutional denoising auto-encoders outperform polynomial fitting and wavelet-based methods when beam effects introduce strong spectral fluctuations, recovering EoR signals with correlation coefficients exceeding 0.92.
These results underscore the potential of deep learning to disentangle complex foreground structures beyond the assumptions of spectral smoothness.

\paragraph{Field-level inference \& HI fraction estimation} represents another frontier where AI offers significant advantages~\citep[see also][]{Acharya02.2026.SKA,Acharya01.2026.SKA}.
Simulation-based inference (SBI; Appendix~\ref{sec:sbi}) can infer astrophysical and cosmological parameters directly from maps or summary statistics, bypassing explicit likelihoods by leveraging amortized neural density estimation (typically through flow-based networks), capturing non-Gaussian information inaccessible to power-spectrum analyses (see Appendices~\ref{sec:norm_flow}~and~\ref{sec:sbi}).
\cite{pietschkeDirectReconstructionReionization2025} demonstrate an application of SBI, EoRFlow, which infers the global neutral hydrogen fraction $\chi_{\text{HI}}(z)$ directly from cylindrically averaged power spectra, enabling piece-wise reconstruction of the reionization timeline; beyond $\chi_{\text{HI}}$, SBI approaches can simultaneously constrain astrophysical parameters such as the ionizing emissivity, mean free path of ionizing photons, and X-ray heating efficiency, or parameters more generally related to star formation~\citep[cf.][]{cerardi_2025_implicit_inference}.
\cite{schosser_2025_optimal_fast} extend this paradigm using invertible neural networks linked to physics-inspired latent representations, achieving rapid posterior estimation with well-calibrated summaries.
Similarly, \cite{tiwariImprovingConstraintsReionization2022} show with Bayesian inference accelerated by ANN emulators and SKA-Low-like noise assumptions that all-shape bi-spectra improve reionization-parameter constraints by factors of 2-4 depending on the reionization stage.
Complementary approaches such as TPCNet~\citep{nguyenTPCNetRepresentationLearning2025} combine convolutional and transformer architectures to infer cold neutral gas fractions and opacity corrections from HI spectra, outperforming CNN baselines in accuracy and robustness.
\cite{choudhuryExtracting21cmPower2022} use ANN to predict 21-cm power spectra from foreground-corrupted synthetic observations and then infer reionization parameters from those predicted spectra; similar works have been carried out using CNNs on 2D power spectra by \cite{laplanteMachineLearningApplied2019, mangenaConstrainingReionizationHistory2020}.
Recently, \cite{salis_2025_exploring_early} performed a comparative study of various neural network architectures for the analysis of 2D power spectra, finding multi-layer perceptron (MLP) mixer and CNN ensembling strategies to provide most flexibility and practicability.
For the direct processing on 3D tomographic image cubes, \cite{neutschInferringAstrophysicsDark2022} introduce a 3D CNN (3D-21cmPIE-Net) to jointly infer Cosmic Dawn and EoR astrophysics alongside warm-dark-matter properties.
Finally, computational costs for likelihood-based inference scale with the number of parameters and novel approaches such as that of~\cite{piras_2024_future_cosmological} can be applied in the context of EoR.
Simulation-based calibration (SBC) provides a complementary diagnostic for any of these inference pipelines: by comparing posterior distributions over inferred parameters against the known simulation inputs, it validates whether posteriors are well-calibrated without requiring tractable likelihoods -- a critical quality-control step before applying SBI methods to real SKA-Low observations~\citep[e.g.,][]{schosser_2025_optimal_fast, semelin_2025_combining_summary, prelogovic_2023_exploring_likelihood}.

\paragraph{Calibration with inductive biases} is essential for mitigating wedge leakage and residual systematics.
Physics-informed neural networks (PINNs; Appendix~\ref{sec:pinns}) and regularized learning strategies incorporate physical constraints into the training objective, reducing spectral structure in gains and beams.
For instance, \cite{tripathiImpactCalibrationPosition2025a} quantify tolerance levels for calibration and antenna-position errors using ANN-Bayesian frameworks, finding that calibration errors must remain below 0.001\% to avoid foreground contamination.
\cite{bianco_2025_deep_learning} recently used binary segmentation map priors of ionized HII regions injected into skip connections of a U-Net to recover the 21-cm signal from SKA-Low's foreground-contaminated observations.
Their framework, SERENEt, achieves up to 90\% accuracy in late-stage signal recovery and over 93\% accuracy in cylindrical power spectra.
Moreover, \cite{korber_2023_pinion} use PINNs for the acceleration of radiative transfer simulations for the forward modelling of cosmic reionization.
In these diverse examples, well studied and calibrated inductive biases improved predictions and enhance the efficiency of signal extraction which is needed to recover the faint 21-cm signal.

\paragraph{Uncertainty-aware detection} is critical for scientific reliability.
Bayesian neural networks (BNNs; Appendix~\ref{sec:bnn}) and deep ensembles provide calibrated posterior predictive distributions, enabling robust parameter inference under limited training data.
This was demonstrated by \cite{mahidaANNBNNInferring2025} which showed that BNN emulators outperform deterministic ANN surrogates for 21-cm power spectrum and bi-spectrum inference, yielding tighter constraints on EoR parameters and highlighting the utility of higher-order statistics.

Current-generation instruments such as LOFAR and MWA lack the sensitivity for direct EoR detection; statistical power-spectrum detection is in principle within reach, but remains hindered by calibration systematics and analysis bottlenecks that leave large fractions of acquired data unanalysed (another situation where AI-driven pipelines can provide significant leverage).
This necessitates reliance on high-fidelity end-to-end simulations for model pretraining.
However, sim-to-real discrepancies, arising from differences in noise characteristics, calibration artefacts, and unmodelled systematics, remain a major challenge.
Domain adaptation techniques, including adversarial training, contrastive representation learning, and fine-tuning on real observations, mitigate these gaps \citep{baron_perez_2025_classification_radio, parker_2024_astroclip, riggi_2024_self_supervised, slijepcevic_2023_radio_galaxy_zoo, margalef-bentabol_2020_detecting_outliers}.
Additionally, domain randomization during simulation, where beam shapes, gain errors, and foreground morphologies are varied, improves generalization.
Combined with uncertainty-aware architectures and physics-informed priors, these strategies enable robust foreground separation and cosmological signal recovery under SKA-scale conditions.
Finally, end-to-end instrument simulations remain foundational utilities for assessing SKA-scale performance of AI-based processing pipelines; the community-wide efforts for the SKAO Science Data Challenge 3a~\citep{bonaldi_2025_square_kilometre} are exemplary demonstrations that current pipelines can recover EoR signals under realistic conditions, though confidence intervals remain underestimated, motivating cross-pipeline comparisons for bias reduction.


\subsection{Radio Galaxies - Detection/Classification}\label{sec:radio_galaxies}

Continuum surveys conducted by \textit{SKA-Mid} (AA4 and beyond) are expected to detect hundreds of millions of radio galaxies spanning diverse redshifts, morphologies, and physical states.
These sources often exhibit complex, multi-component structures, such as lobes, jets, bent tails, and diffuse haloes, that challenge traditional source-finding and classification pipelines~\citep[see also][]{Hale01.2026.SKA,Sasmal01.2026.SKA}.

Established automated tools such as PyBDSF~\citep{mohan_2015_pybdsf} and Aegean~\citep{hancock_2012_aegean_1, hancock_2018_aegean_2} have proven reliable workhorses for detecting and cataloguing compact and moderately extended sources in precursor surveys, performing well on the well-characterized, high-SNR cases that dominate source counts~\citep[see also][]{SabyasachiPal01.2026.SKA}.
However, they rely on parametric fitting and fixed morphological assumptions that break down precisely where the science is most demanding: associating disjoint multi-component sources, distinguishing faint diffuse emission from artefacts, classifying rare or ambiguous morphologies, and generalizing robustly across heterogeneous noise and uv-coverage conditions.
Manual inspection and rule-based systems are infeasible at SKA scales, particularly in low SNR regimes and for rare or poorly labelled morphologies.
Consequently, automated, morphology-aware approaches capable of learning flexible, data-driven representations are essential.

\paragraph{Morphology-aware detection and segmentation} has become a central application of deep learning in radio astronomy~\citep[see also][]{Prandoni01.2026.SKA,Hardcastle01.2026.SKA}.
Architectures, such as YOLO (Appendix~\ref{sec:yolo}), Mask R-CNN (Appendix~\ref{sec:mask_rcnn}), SAM (Appendix~\ref{sec:sam}), U-Net (Appendix~\ref{sec:unet}), and Vision Transformers (ViTs; Appendix~\ref{sec:transformer}) enable pixel-level segmentation, object detection (Appendix~\ref{sec:object_detection_segmentation}), and classification of extended structures, outperforming classical algorithms in associating disjoint emission regions and identifying multi-lobed sources.

\cite{riggiAstronomicalSourceDetection2023, riggi_2024_detection_classification} present a Mask R-CNN--based source finder for SKA-precursor continuum surveys like ASKAP EMU, reducing false detections in the Galactic plane and improving association of fragmented emission islands; similar studies have been carried out by \cite{burke_2019_deblending_classifying, mostert_2022_radio_source_component, zhang_2022_automatic_detection}.
\cite{sortino_2023_radio_astronomical} provides a comparative study of object detection models such as Mask R-CNN, Detectron2, DETR, and EfficientDet, finding promising results of above 90\% accuracy for radio astronomical images.
The SKAO Data Challenge 1 \citep{bonaldi_2018_square_kilometre} primarily focused radio source finding, for which \cite{cornu_2024_yolo_cianna} developed YOLO-CIANNA, a customized deep-learning object detector; the model was later applied in the SDC2, ranking at the top of the scoring board \citep{cornu_2026_yolo_cianna}.
It outperforms previous methods in both detection purity and the number of sources found, while maintaining high computational efficiency and strong characterization accuracy.
Similarly, \cite{tolleyWaveletScatteringNetworks2024} investigates Fanaroff–Riley (FR) morphology classification using scattering transforms, showing robustness under noise injection and training-set truncation, and achieving performance competitive with CNN-based classifiers.

\paragraph{Scalable survey analysis and cross-validation} requires AI pipelines integrated into SKA workflows for wide-field source detection, classification, and multi-wavelength cross-identification.
Transfer learning and contrastive learning approaches (see Appendices~\ref{sec:supervised_unsupervised} and \ref{sec:self_supervised}) enhance robustness across instruments and surveys.
\cite{lastufkaSelfSupervisedLearningMeerKAT2024} show that self-supervised learning can be trained directly on wide-field MGCLS MeerKAT continuum crops without single-galaxy cutouts, using DINO with ResNet50 and ViT backbones to learn reusable representations for SKA-era surveys.
Their SSL models reach state-of-the-art FRI/FRII classification and much better compact-source count prediction, while requiring only $\sim$20k source-rich crops and yielding fine-tuned multi-class performance comparable to supervised and natural-image foundation models. 

\paragraph{Vision foundation models} (VFMs; Appendix~\ref{sec:vfms}), pretrained on massive natural-image datasets using self-supervised learning (Appendix \ref{sec:self_supervised}), offer unprecedented feature extraction capabilities for radio astronomy.
Models such as CLIP, SAM, and DINO can be adapted for tasks like FRI/FRII classification, AGN identification, or source count estimation, even in low-label regimes, reducing the need for extensive manual annotation in long-term citizen science projects~\citep[see also][]{Hota01.2026.SKA}.
Their high-level semantics and low-level patterns captured during training with natural images provide a starting point in the radio data domain, reducing the computational effort (and data volumes) to learn those aspects from scratch.
\cite{lastufka_2025_examining_vision} evaluates how vision foundation models can be adapted for optical and radio astronomy tasks, finding that these models improve classification accuracy for optical galaxy images and perform well in radio object detection, but struggle with radio galaxy classification.
Going even a step further than VFMs, \cite{riggi_2025_llava} study fine-tuning of a vision-language model (LLaVA) on a large set of radio images and captions.
They achieve about 30\% F1-score gains in extended source detection compared to base models, but still under-performing vision-only classifiers and showing a 20\% drop on general multimodal tasks.
This work highlights both the promise and current limitations of such models for radio astronomy, including the need for better multimodal alignment and higher-quality datasets.
Similarly, \cite{drozdova_2025_radio_astronomy} evaluate various vision–language models for classifying radio galaxy morphologies, finding that prompt-based approaches can perform well even without astronomy-specific training, but their outputs are highly sensitive to prompt variations.

\paragraph{Data augmentation} using generative models, GANs, VAEs, and diffusion models (see Appendices \ref{sec:gan}, \ref{sec:vae}, and \ref{sec:diffusion}), further improves classifier generalization under limited data conditions.
Synthetic radio galaxy images generated by these models replicate realistic morphologies, enabling robust training for rare classes and mitigating class imbalance.

Class-imbalance is a common problem among galaxy detection and classification tasks and can lead to biases towards the well-represented classes.
Appropriate sampling strategies, synthetic augmentation, and anomaly-aware training can alleviate such problems.
Due to the inherently high data compression/reduction of tasks such as physical parameter inference or morphological classification (from images to a few scalars or tokens), it is difficult to retrace single decisions of models and retain scientific trust.
Thus, it is important to validate results using interpretability techniques (Appendix~\ref{sec:xai}), such as feature attribution (through saliency maps via Grad-CAM or SmoothGrad), prototype-based explanations, or counterfactual analysis.
Simulation-based calibration can validate whether posterior distributions produced by probabilistic inference methods are well-calibrated through comparison against simulated parameters.
This provides a diagnostic tool without requiring ground-truth likelihoods and ultimately increases trustworthiness of AI outputs.
Finally, the adoption of MLOps best practices prevents the occurrence of pain points (operational failure modes such as silent model drift, training-serving skew, and data version mismatch), contributes to the reproducibility of models, and provides means to monitor model drift and performance after deployment.


\subsection{Pulsar \& Transient Science}\label{sec:pulsar_transient_science}

The SKA is set to revolutionize time-domain radio astronomy by enabling the discovery of fast radio bursts (FRBs), exotic pulsar systems, and other transient phenomena across a vast range of timescales and astrophysical regimes~\citep[see also][]{Joshi01.2026.SKA,Miller-Jones01.2026.SKA,Keane01.2026.SKA,Shannon01.2026.SKA}.
They are primary science drivers for \textit{SKA-Mid} (AA4), which provides the sensitivity, frequency coverage, and timing precision required for pulsar timing arrays, FRB localization, and pulsar discovery. \textit{SKA-Low} contributes a more limited but complementary role, enabling searches for long-period pulsars and low-frequency transient phenomena in a regime inaccessible to SKA-Mid~\citep[see also][]{Qiu01.2026.SKA}.
In general, the scale and cadence of SKA data, spanning petabyte-scale archives and real-time beam-formed streams, render traditional manual and heuristic approaches infeasible.
Classical pipelines struggle with RFI, rare-event imbalance, and latency constraints, particularly in the context of real-time follow-up and precision timing.
To fully exploit SKA's sensitivity and temporal resolution, automated, adaptive, and scalable analysis frameworks are required.

\paragraph{Candidate triage and prioritization} is a critical bottleneck in transient discovery.
Deep learning models, including CNNs and transformers (see Appendices~\ref{sec:cnn} and \ref{sec:transformer}), can automate the ranking and filtering of transient candidates from high-volume data streams, reducing expert workload and enabling dynamic prioritization of events based on scientific relevance~\citep[cf.][]{sravan_2022_autonomous_real_time, agarwal_2020_fetch, connor_2018_applying_deep}.
These systems adapt to changing noise environments, and optimize telescope time allocation for follow-up resources~\citep[see also][]{GemmaAnderson01.2026.SKA}.

\paragraph{Anomaly detection} for rare events leverages unsupervised and semi-supervised learning to identify low-likelihood signals without laboriously labelled datasets~\citep[see also][]{AlexAndersson02.2026.SKA,AlexAndersson01.2026.SKA}.
Techniques such as one-class classifiers, VAEs, and deep generative models (see Appendices \ref{sec:vae} and \ref{sec:anomaly_detection}) have shown promise in detecting novel transients~\citep[e.g.,][]{kuiper_2025_representation_learning}.
Active learning strategies further refine these models by incorporating minimal human feedback.
\cite{anderssonFindingRadioTransients2025} benchmark anomaly-detection algorithms on MeerKAT light curves, demonstrating that ranking by anomaly score recovers over half of known transients within the top 10\% of candidates, while active learning improves recall by up to 20 percentage points when supplemented with 2\% labelled data.

\paragraph{Real-time pulsar search} with edge processing is indispensable for SKA-scale operations.
Lightweight CNN-based architectures (Appendix~\ref{sec:cnn}) can be directly deployed on telescope back-ends or FPGA-based edge devices, enabling on-the-fly analysis during acquisition.
This approach mitigates the risk of losing weak or short-lived signals during aggressive data compression and circumvents the impracticality of archival reprocessing.
\cite{fuImprovingPulsarSearch2025} present a deep-learning–assisted pulsar-search pipeline developed and validated on current-generation telescopes (FAST, GBT, Arecibo, and MWA) that ranks folded ``snapshot'' candidates from de-dispersed time series, so the costly \textit{folding stage}, in which candidate periods are used to coherently fold thousands of time-series samples into a pulse profile for visual or automated inspection, is invoked far less often in next-generation surveys.
The pipeline delivers $\sim$10--60 fold speed-ups while retaining nearly 100\% recall, demonstrating strong cross-instrument generalization and direct methodological relevance for SKA pulsar surveys.

\paragraph{Precision enhancement and noise reduction} represents another domain where AI excels, and the techniques discussed here are broadly applicable across the scientific areas covered in this chapter.
Conditional generative models, such as GANs, diffusion models, and flow-matching architectures (Appendix \ref{sec:cgan}, \ref{sec:diffusion}, or \ref{sec:flow_matching}), can denoise and super-resolve pulse profiles, improving timing accuracy for applications like pulsar timing arrays in gravitational wave detection~\citep{li_2021_detection_radio}.
These methods preserve temporal resolution while suppressing noise and instrumental artefacts.
For instance, \cite{zhang_2024_rfi_drunet} reframe RFI mitigation in pulsar observations (for LOFAR/NenuFAR) as a joint detection-and-restoration problem on the full time-frequency dynamic spectrum.

\paragraph{Distributed learning} frameworks, particularly federated learning (FL; Appendix~\ref{sec:federated}), enable collaborative model training across observatories without centralizing raw data, reducing bandwidth demands and enhancing privacy.
This paradigm is particularly relevant for global transient detection and large-scale pulsar surveys.
Unfortunately, there is a significant research gap in the application:
while federated model training has remained entirely unexplored in the context of exascale radio interferometry, works by \cite{scaife_2020_big_telescope} and \cite{wang_2025_decentralized_framework} share some architectural similarities with FL.

\paragraph{Autonomous decision-making and scheduling} can be optimized using reinforcement learning (RL; Appendix~\ref{sec:reinforcement}).
RL agents learn policies that balance scientific yield with operational constraints, dynamically allocating resources for follow-up observations and survey planning.
Recent work also demonstrates the use of RL to automate calibration workflows in radio interferometry for current-generation instruments \citep[cf.][]{kirk_2024_reinforcement_learning}, with the explicit goal of developing transferable approaches for SK-scale autonomous operations.
Such calibration agents can generalize across diverse observations, minimizing the need for manual intervention and enabling ``smart calibration'' of radio telescopes \citep{yatawatta_2023_hint_assisted}; so far, the SKAO has not considered autonomous scheduling and adaptive calibration as key operational requirements.

Despite these advances, AI-driven systems must be designed with safeguards to ensure scientific reliability, especially those with a high degree of autonomy.
Continuous monitoring and retraining strategies are essential to avoid model drift and maintain performance under evolving noise and signal conditions.
Diverse training datasets (in particular when data is synthetic) help prevent bias toward known classes, while anomaly-aware architectures improve robustness.
Moreover, explainable AI techniques (see Appendix~\ref{sec:xai}), such as saliency maps, integrated gradients, and counterfactual analysis, enhance transparency and increase trust in results during human validation~\citep[cf.][]{denzel_2024_towards_certification}.
Furthermore, MLOps practices (Appendix~\ref{sec:mlops}) aid in seamless integration of the aforementioned explainable AI techniques, and monitoring models after deployment \citep{billeter_2024_mlops_enabler}.


\subsection{Cosmic Magnetism}\label{sec:cosmic_magnetism}

Understanding cosmic magnetism is central to explaining galaxy evolution, star formation, and cosmic ray transport \citep[cf.][]{carretti_2025_radio_observations}.
Observables such as Zeeman splitting and Faraday rotation encode magnetic field strength and topology, yet these signals are intrinsically faint and easily masked by instrumental artefacts and foreground contamination~\citep[see also][]{Vernstrom01.2026.SKA,Robishaw01.2026.SKA,Bourke01.2026.SKA}.
SKA-AA$\ast$ will deliver an unprecedented number of Zeeman measurements in star-forming regions, and provide invaluable insights into the strength and structure of magnetic fields in the interstellar medium (ISM), however, disentangling these faint signals from Stokes V spectra remains challenging~\citep[cf.][]{pattle_2022_magnetic_fields}; AA$\ast$ is the only configuration with enough sensitivity to detect magnetic field strengths of the order of a few $\mu$G~\citep{heald_2020_magnetism_science}.
Similarly, extracting extra-galactic Faraday rotation signals from rotation measure (RM) catalogues requires separating Galactic contributions and measurement uncertainties, a task that classical approaches struggle to achieve at SKA scales \citep[cf.][]{takahashi_2023_introduction_to}.

\paragraph{Radio polarimetry} is the first step in the study of cosmic magnetism.
High-fidelity imaging in the Stokes Q and U polarization parameters are essential components in the study of cosmic magnetism.
\cite{wang_2023_polarrec} address the fundamental data representation in deep learning models for this first step in the data processing pipeline, developing and validating their approach on current interferometric datasets with direct applicability to \textit{SKA-Mid} polarimetric imaging.
While traditional CNNs operate on Cartesian grids, their PolarRec framework operates in polar coordinates along elliptical tracks in the uv-plane, and demonstrates superior performance using transformer encoders, reducing spurious magnetic signal artefacts.

\paragraph{Zeeman line profile recovery} can benefit substantially from machine learning.
Models trained on synthetic and real Stokes V spectra learn non-linear mappings that isolate weak Zeeman signatures from instrumental distortions, enabling robust magnetic field estimation in molecular clouds.
Diffusion models, in particular, offer flexibility for recovering magnetic field strength from multiple observables, including column density, dust polarization angles, and line-of-sight velocity dispersions (Appendix~\ref{sec:diffusion}).
\cite{xu_2025_exploring_magnetic} use such diffusion models trained on synthetic ISM observables to estimate magnetic field strength detectable through SKA-Mid AA$\ast$.
With a 3-channel diffusion model (using column density, polarization angles, and velocity dispersion) they outperform both traditional power-law fitting and Davis–Chandrasekhar–Fermi (DCF) methods, especially on previously unseen data.
The classical DCF method overestimated field strength by about an order of magnitude, while the 3-channel diffusion demonstrated superior precision and robustness.

\paragraph{Faraday rotation disentanglement} is another critical application where AI can optimize bottlenecks.
Extracting detailed RM structures from radio data under complex Faraday dispersion and noise conditions can be formulated as an inverse problem solvable by generative models~\citep[see also][]{Carcamo01.2026.SKA,OSullivan01.2026.SKA,Loi01.2026.SKA}.
Models consisting of simple U-Nets (Appendix~\ref{sec:unet}), conditional GANs (Appendix~\ref{sec:cgan}), or diffusion models (Appendix~\ref{sec:diffusion}) have demonstrated superior performance compared to traditional algorithms like RMCLEAN~\citep{gustafssonSemiSupervisedRotationMeasure2024}, with studies conducted primarily on MeerKAT data as a direct SKA-Mid precursor~\citep[e.g.][]{ndiritu_2021_gaussian_process, carcamo_2022_cs_romer}.
\cite{gustafssonSemiSupervisedRotationMeasure2024} present a semi-supervised deep-learning model for Faraday spectrum deconvolution, achieving accurate recovery of dispersion functions across a broad RM range while scaling efficiently to large MeerKAT datasets.
This approach reveals intricate magnetic structures in cluster relics and AGN, highlighting the potential of AI for high-resolution magnetism studies.

\paragraph{3D mapping of magnetic fields} represents a frontier where AI-driven methods outperform classical techniques~\citep[see also][]{Tahani01.2026.SKA,Sun01.2026.SKA,Ma01.2026.SKA}.
Conditional generative models (Appendix~\ref{sec:cond_generation}) trained on magneto-hydrodynamic simulations can predict 3D field orientation, strength, and turbulence parameters from complex, noisy observational proxies such as HI emission, synchrotron maps. These models surpass velocity gradient and equi-partition methods in accuracy and robustness, even under low SNR conditions.
\cite{hu_2024_probing_three, hu_2025_probing_three} demonstrate this by training a CNNs and ViTs on synthetic data to robustly extract the plane-of-the-sky orientation, inclination angle, and magnetization from synchrotron observations.
Similarly, \cite{hu_2025_characterizing_magnetic} present a deep learning framework based on conditional ResNets to predict the full 3D magnetic field structure and turbulence properties in HI clouds from spectroscopic observations.
Moreover, \cite{denzel_2025_galactic_alchemy} demonstrates magnetic field strength recovery from multiple other quantities such as gas column density with conditional GANs and diffusion models trained on magneto-hydrodynamics simulations. While their reconstructions are technically only 2D, their approach is straight-forward to adapt to full 3D reconstruction of magnetic fields within galaxies.

Similar to many other scientific fields, counting of events with time-independent probability follow a Poisson distribution, which means the noise variance is proportional to square-root of the signal and can generally be reduced with more acquisition time.
When longer integration times are impractical or signals are still intrinsically close to the SNR level, it has been demonstrated that model training on samples artificially enhanced with simple noise simulations can improve real-world performance \citep[cf.][]{oppliger_2024_weak_signal}.
While CNNs (Appendix~\ref{sec:cnn}) and ViT (Appendix~\ref{sec:transformer}) maintain high accuracy even with high noise and missing data, making them suitable for low signal-to-noise ratio (SNR) observations, their predictions are not immediately transparent.
\cite{peek_2019_do_androids} investigate the interpretability of such models, showing that learned features can be meaningfully linked to physical properties such as turbulence anisotropy and morphological curvature, thereby partially alleviating black-box concerns.


\subsection{Gravitational Lensing}\label{sec:gravitational_lensing}

Gravitational lensing, particularly in the strong regime, is the key probe of dark matter and its distribution on small scales.
Low-mass dark matter haloes are of special interest because their population mass function allows direct testing of competing dark matter models (cold versus warm dark matter).
Lensed radio sources provide constraints on dark matter properties through magnification and differential time delay measurements; \textit{SKA-Mid} (AA4) is poised to discover thousands of such systems \citep[cf.][and references therein]{mckean_2015_strong_gravitational}, rendering manual identification and modelling infeasible at survey scale.

Unlike optical observations, radio lensing analysis faces unique challenges: the morphological characteristics of radio sources are harder to distinguish from lensing features due to typically lower angular resolution and SNR, and radio emission traces gas rather than stellar continuum, complicating direct transfer of optical pipelines.
Consequently, AI-based work on radio gravitational lensing remains scarce compared to the optical domain, partly due to the complexity of interferometric visibility-space data and the comparatively greater availability of wide-field optical imaging.

Nevertheless, the AI architectures and inference frameworks developed extensively in the optical regime provide methodological templates for radio application, and we review both here, emphasising the adaptations needed for interferometric data.
Beyond methodology transfer, the combination of radio and optical observations of the same lensing systems offers qualitatively stronger science: optical data constrain the lens and source light distributions, while radio interferometric data provide independent astrometric constraints, probe gas-phase source morphology invisible in optical bands, and are free of dust obscuration -- together breaking degeneracies such as the mass-sheet degeneracy that neither waveband can resolve alone. This argument for cross-spectrum synergy is not unique to gravitational lensing (although more apparent due to the data scarceness); it applies broadly across SKA science, where joint radio--optical--infrared analyses can disentangle physical components, validate detections, and tighten parameter constraints in ways that no single waveband can achieve independently~\citep[cf.][]{prandoni_2015_revealing_physics}.

\paragraph{Automated lens finding and modelling} are both well developed in the optical regime and provide the template for radio application.
For \textit{identification}, CNNs substantially outperform traditional methods (see Appendix~\ref{sec:cnn}), with single-stage object detectors such as YOLO offering the throughput required at survey scale \citep[e.g.][]{liu_2025_yolox_ls}, and unsupervised/anomaly-detection approaches (see Appendix~\ref{sec:supervised_unsupervised} and~\ref{sec:anomaly_detection}) -- notably auto-encoders -- exploited for the rare-event regime \citep[e.g.][]{cheng_2020_identifying_strong, etsebeth_2024_astronomaly_at_scale}.
The one explicitly radio-interferometric demonstration to date is \cite{rezaeiMachineLearningBased2022}, who tailored CNN-based lens finding to LOFAR images and recovered $95.3\%$ of true positives at a $0.008\%$ false-positive rate, implying that most galaxy-scale lenses in LoTSS should be detectable.
For \textit{modelling}, Bayesian neural networks and simulation-based inference stochastically infer parametric lens-model parameters (Appendix~\ref{sec:sbi}; e.g.\ \citealp{legin_2021_simulation_based}), while parameter-free reconstructions using Recurrent Inference Machines (RIM) properly explore the degenerate solution space \citep{morningstar_2018_analyzing_interferometric, morningstar_2019_data_driven}; through explicit integration of gradient log-likelihoods, probabilistic methods such as RIM or diffusion models can enforce physical priors and observational constraints to suppress implausible solutions \citep[cf.][]{remy_2023_probabilistic_mass, rustigIntroducingLensCharmCharming2024}.
Crucially for SKA, differentiable lens-modelling frameworks such as PyAutoLens~\citep{nightingale_2018_autolens} are already compatible with radio interferometric and visibility data and can serve as a template for radio-adapted, AI-based modelling.

Due to the inherently ill-posed nature of the gravitational lensing problem the identification and modelling of gravitational lensing require precise physical constraints and should be interpretable to assess model plausibility. This is in direct conflict with the ``black box'' nature of deep learning.
The rarity of the phenomenon also leads to another critical risk of supplementing or replacing actual observational datasets with simulations and therefore widening the sim-to-real domain gap for neural networks.
Furthermore, morphological mimics pose a significant challenge; intrinsic radio structures such as bent-tail galaxies or chance alignments of star-forming clumps can closely resemble lensing arcs, potentially contaminating catalogues in the absence of redshift information.
There is also the risk of training set bias, where supervised models favour systems with large Einstein radii or high SNR ratios, effectively blinding the survey to lower-mass lenses or compact configurations that are crucial for dark matter studies.

These risks can be mitigated using a variety of strategies: physics-informed architectures that explicitly incorporate the radio interferometric measurement equation in to the loss function or architecture (such as in RIM) ensure that any reconstructed lens model adheres to mathematical consistency with sampled visibilities, thereby reducing artefacts.
Optical-to-radio domain and sim-to-real gaps can be mitigated using unsupervised domain adaptation techniques~\citep{swierc_2024_domain_adaptive}. They allow networks to learn feature representations that are invariant to the differences between simulations and real data and have been shown to improve inference accuracy by orders of magnitude when applied to noise properties that differ from the training set.
To address computational bottlenecks and one-shot reliability issues when dealing with large datasets, a hierarchical inference pipeline could be considered: in a first phase smaller, fast networks such as CNNs are used for initial candidate discovery and artefact rejection \citep[as in][]{hezaveh_2017_fast_automated}, calibrated for high coverage and low false-positive rate, followed by a more rigorous, physics-informed approach such as SBI for final parameter estimation of high-probability candidates.
Finally, considering the comparably large number of lenses already detected with other instruments such as Euclid or JWST, a multi-wavelength approach provides validation opportunities and could provide even stronger constraints for lens models to break/alleviate the mass-sheet degeneracy. Moreover, since radio emission and optical emission trace different physical components of galaxies, a true lens candidate should exhibit consistent astrometry across bands, while mimics more often do not. Many deep learning frameworks provide the flexibility to integrate multiple data domains for lens candidate analysis.


\section{Conclusion}

\begin{figure}
  \centering
  \includegraphics[width=0.99\textwidth]{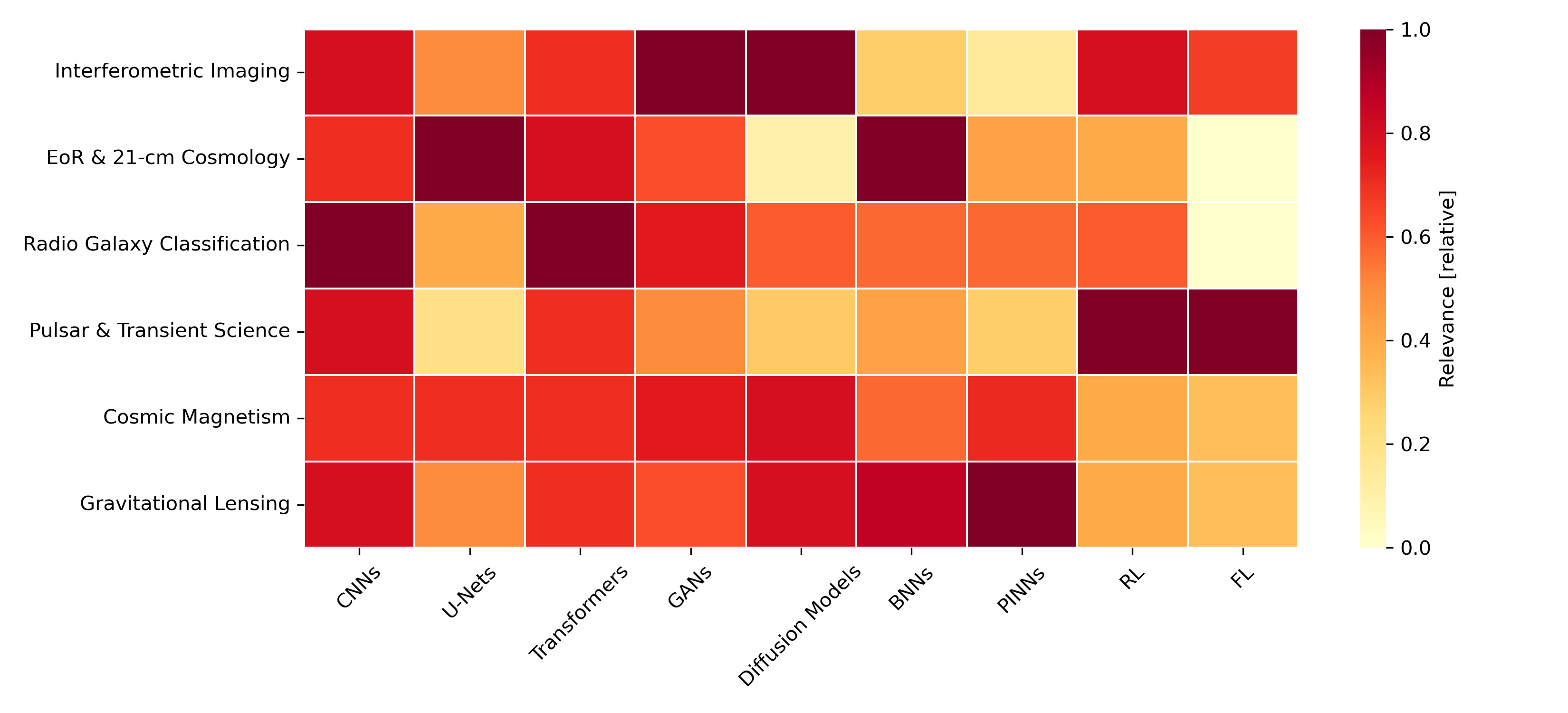}
  \caption{
    A semantic cross correlation heatmap, qualitatively illustrating the relative importance of AI-based methods in SKA problem areas. Note that this figure is derived from the references discussed in this chapter and reflects an indicative rather than exhaustive selection of topics. However, the frequency of each AI-based method is representative when compared against matched papers from arXiv.
  }\label{fig:heatmap}
\end{figure}

The Square Kilometre Array Observatory (SKAO) represents a paradigm shift in radio astronomy, not only in terms of sensitivity and survey speed but also in the scale and complexity of its data ecosystem.
This chapter has outlined how Artificial Intelligence (AI) is uniquely positioned to address these challenges, enabling real-time operations, accelerating scientific workflows, and unlocking new discovery modes across cosmology, galaxy evolution, and time-domain astrophysics.
From deep learning–based interferometric imaging and generative sky simulations to uncertainty-aware inference and physics-informed architectures, AI has demonstrated transformative potential in domains where classical methods fail under SKA-scale conditions (cf. Figure~\ref{fig:heatmap}).

Yet, the integration of AI into SKA science is not without caveats.
The opacity of deep learning models, risks of hallucinated structures, and domain gaps between simulations and observations underscore the need for rigorous validation and interpretability.
Embedding physical constraints, adopting Bayesian frameworks for uncertainty quantification, and implementing robust MLOps practices -- including data versioning, model registries, and continuous performance monitoring across the distributed SKA data infrastructure -- will be essential to maintain scientific integrity at AA4 scale and beyond.
These safeguards must evolve alongside increasingly autonomous systems, ensuring that AI remains a tool for discovery rather than a source of bias or error.

Looking ahead, several research frontiers merit attention.
First, scalable simulation-based inference \citep{cranmer2020frontier, cerardi_2025_implicit_inference} will become indispensable for extracting non-Gaussian information from SKA datasets, bridging theory and observation through amortized neural density estimation.
Second, foundation models and self-supervised learning \citep{parker_2024_astroclip, lastufka_2025_examining_vision} promise to reduce reliance on labelled data, enabling cross-survey generalization and multimodal integration with optical, infrared, and gravitational-wave observations.
Third, distributed and federated learning \citep{kairouz2021advancesopenproblemsfederated} will support collaborative science across global SKA nodes without centralizing raw data, addressing bandwidth constraints.
It should be noted, however, that the SKAO data management framework is currently built around a network of centralized SKA Regional Centres (SRCs), and federated learning is not part of the SKAO operations road map at this stage \citep{srcnet_docs_2025, srcnet_v01_2025}. The relevance of federated approaches will depend on how data access policies and SRC governance evolve, particularly with respect to proprietary data periods and the eventual opening of archival datasets to the broader community.
Finally, reinforcement learning and adaptive scheduling \citep{kirk_2024_reinforcement_learning, yatawatta_2023_hint_assisted} could optimize telescope operations in real time, balancing scientific yield against resource limitations.

The SKA era will not merely scale existing workflows; it will redefine the scientific method in radio astronomy.
AI has already contributed to advances across the board in SKAO science.
From the acceleration of forward models, over the automation of real-time analyses, to the dynamic compression of unprecedented data volumes, AI provides flexible and efficient solutions that lighten the load on computational pipelines and field experts alike.
The evidence suggests that AI is no longer merely an optimization tool but a prerequisite for unlocking the physics of the Universe in the radio spectrum, offering necessary breakthroughs in scalability, fidelity, and the serendipitous discovery of ``unknown unknowns''.

\appendix

\section{AI Methods Glossary}\label{sec:ai_methods}

The following provides a concise reference summary of the main AI concepts and methods cited in Section~\ref{sec:scientific_problems}, with a focus on learning paradigms, common tasks and learning objectives, neural network architectures, as well as deep generative models.
Where applicable, the SKA-relevant use case for each method is noted; of course, this application area relevance is non-exclusive.

\subsection{Learning Paradigms and Development Patterns}

\subsubsection{Deep Learning}
Deep learning is a sub-field of machine learning that trains multi-layer artificial neural networks to learn hierarchical representations of data, enabling end-to-end learning of complex tasks such as image recognition, speech recognition, machine translation, and scientific discovery \citep{lecun2015deep,Goodfellow-et-al-2016}. It relies on large datasets, gradient-based optimization (back-propagation), and architectural innovations (convolutional, recurrent, transformer-based networks) to automatically extract features at multiple levels of abstraction and has driven many recent advances in AI.

\textit{SKA relevance}: the foundational paradigm underlying all deep learning methods applied across the scientific areas in Section~\ref{sec:scientific_problems}.

\subsubsection{Supervised and unsupervised Learning}\label{sec:supervised_unsupervised}
Supervised learning trains models on labelled examples (input–output pairs) so they learn to map inputs to known targets, optimizing a loss that measures prediction error; it excels at tasks like classification and regression when annotated data are available (e.g., image labels, paired translations). Unsupervised learning discovers structure from unlabelled data by modelling data distributions, finding clusters, or learning useful representations (e.g., clustering, density estimation, auto-encoders), and is useful when labels are scarce. Semi-supervised and self-supervised methods bridge the two by leveraging large unlabelled datasets plus limited labels to improve performance \citep{mitchell1997machine}.

\textit{SKA relevance}: supervised learning underpins source detection and classification pipelines; unsupervised and semi-supervised approaches are critical where labelled radio data is scarce, as is common for rare morphologies (Section~\ref{sec:radio_galaxies}) and EoR signal extraction (Section~\ref{sec:eor}).

\subsubsection{Self-supervised learning (SSL)}\label{sec:self_supervised}
Self-supervised learning trains models using automatically generated supervision from the data itself by defining pretext tasks (e.g., predicting masked inputs, contrasting augmented views, or solving context/order prediction) so models learn rich representations without manual labels; these learned features can be fine-tuned or transferred to downstream tasks and have driven major advances in vision, language, and multimodal learning (e.g., masked language modelling, contrastive learning, and masked image modelling).

\textit{SKA relevance}: SSL enables learning from the vast unlabelled datasets that SKA-era surveys will produce, reducing annotation bottlenecks for radio galaxy classification and continuum surveys (Section~\ref{sec:radio_galaxies}).

\subsubsection{Reinforcement learning (RL)}\label{sec:reinforcement}
Reinforcement learning (RL) studies how an agent learns to make sequential decisions by interacting with an environment to maximize cumulative reward: the agent observes states, takes actions according to a policy, and updates that policy using feedback (rewards and state transitions) to improve long-term performance; RL spans model-free methods (Q-learning, policy gradients, actor-critic), model-based methods, and modern deep RL that combines neural function approximators with classic algorithms to solve high-dimensional control, games, and robotics problems \citep{sutton1998reinforcement,mnih2015human,murphy2024reinforcementlearningoverview}.

\textit{SKA relevance}: RL is being explored for autonomous telescope scheduling, calibration workflow optimisation, and adaptive follow-up for transient events (Section~\ref{sec:pulsar_transient_science}).

\subsubsection{Federated learning}\label{sec:federated}
Federated learning is a distributed machine-learning paradigm where model training is performed across many client devices (e.g., mobile phones, edge sensors) that keep raw data locally and only send model updates (gradients or weights) to a coordinating server for aggregation, preserving data locality and reducing the need to centralize sensitive datasets \citep{mcmahan2016communicationefficientlearningdeepnetworks,kairouz2021advancesopenproblemsfederated}.

\textit{SKA relevance}: federated learning is a candidate paradigm for training shared models across distributed SKA data centres in South Africa and Australia without centralizing petabyte-scale raw data. So far, this is not on the SKAO operations road-map.


\subsubsection{Machine Learning Operations (MLOps)}\label{sec:mlops}
MLOps (Machine Learning Operations) is the set of practices, tools, and cultural patterns that combine software engineering, DevOps, and data science to reliably deploy, monitor, and maintain machine learning systems in production; it covers the full model life-cycle--data versioning and pipelines, reproducible training, model validation, CI/CD for models, scalable serving, monitoring for data and model drift, and governance (including reproducibility, explainability, and compliance)--with aims to reduce time-to-production, improve reliability, and enable continuous improvement of ML services \citep{sculley2015,Breck2017,huyen2022designing}.

\textit{SKA relevance}: MLOps frameworks are essential for deploying, monitoring, and maintaining AI pipelines across the full SKAO data life-cycle, from real-time telescope back-ends to science-archive post-processing.

\subsubsection{Explainable/interpretable AI/ML}\label{sec:xai}
Interpretable and explainable AI/ML \citep{guidotti2018survey,molnar2020interpretable} concerns methods and practices that make model behaviour, predictions, and learned representations understandable to humans--either by using inherently interpretable models (rules, generalized additive models, sparse linear models) or by applying post-hoc explanation techniques (feature attributions, saliency maps, counterfactuals, surrogate models)--so stakeholders can trust, validate, debug, and meet regulatory or ethical requirements. In practice, explainable and interpretable AI are often used interchangeably (we follow this practice throughout this chapter); however, there is a noteworthy distinction. \textit{Interpretable AI} refers to models that are inherently transparent by design -- you can inspect their structure directly to understand why a prediction was made. Simple examples include linear regression, decision trees, or sparse rule-based systems: the model itself is the explanation. Explainable AI refers to post-hoc techniques applied to models that are not inherently transparent.

\textit{SKA relevance}: explainability techniques such as saliency maps and counterfactual analysis are used to validate source detection, morphology classification (Section~\ref{sec:radio_galaxies}), and EoR inference models (Section~\ref{sec:eor}), building scientific trust in AI outputs.


\subsection{Tasks and learning objectives}\label{sec:tasks}

\subsubsection{Image Classification}
Image classification is a fundamental task in computer vision where an AI model, typically a Convolutional Neural Network (CNN), analyses an entire image's pixel data and assigns it a single, predefined categorical label based on its visual content \citep{toennies2024introduction}. Unlike object detection, which localizes multiple objects, image classification provides a high-level summary of the image's primary subject, a capability often benchmarked using large-scale datasets like ImageNet \citep{deng2009imagenet}.

\textit{SKA relevance}: applied to radio galaxy morphology classification (FRI/FRII and beyond) and pulsar candidate scoring in SKA-Mid and SKA-Low surveys (Section~\ref{sec:radio_galaxies}).

\subsubsection{Object Detection \& Image Segmentation}\label{sec:object_detection_segmentation}

Object detection is a fundamental task within computer vision that enables simultaneous identification as well as localization (by means estimating a bounding box) of objects within images. By contrast, image segmentation is a more granular and sophisticated task than object detection, as it involves classifying every single pixel in an image to belong to a specific object class or background. This process effectively creates precise boundaries or masks for objects, providing a pixel-level understanding of the scene. Popular architectures include Mask R-CNN (Appendix~\ref{sec:mask_rcnn}), YOLO (Appendix~\ref{sec:yolo}) or SAM (Appendix~\ref{sec:sam}).

\textit{SKA relevance}: core task for radio source finding and multi-component association in SKA-Mid continuum surveys; also used for RFI segmentation across both telescopes (Sections \ref{sec:radio_interferometric_imaging} and \ref{sec:radio_galaxies}).

\subsubsection{Vision foundation models (VFMs)}\label{sec:vfms}
Vision foundation models are large, pretrained visual models learned from massive, diverse image (and often image–text) datasets using self-supervised or weakly supervised objectives; they produce versatile, transferable representations that enable a wide range of downstream tasks (classification, detection, segmentation, image captioning, and multimodal reasoning) via fine-tuning or prompting, and have spurred architectures combining vision transformers, contrastive and masked modelling, and multimodal alignment, e.g., CLIP \citep{radford2021learningtransferablevisualmodels} or DINO \citep{caron2021emergingpropertiesselfsupervisedvision}.

\textit{SKA relevance}: VFMs pretrained on optical data are being fine-tuned for radio source classification and detection (Section~\ref{sec:radio_galaxies}), offering strong baselines in low-label regimes relevant for SKA-era citizen science and automated pipelines.

\subsubsection{Anomaly Detection}\label{sec:anomaly_detection}
Anomaly detection identifies patterns in data that deviate from expected behaviour (outliers, novelties, or rare events) to flag errors, fraud, faults, or unusual phenomena; methods range from simple statistical thresholds and distance- or density-based techniques (e.g., k-NN, isolation forest, local outlier factor) to reconstruction- or representation-based approaches using auto-encoders, one-class support vector machines, and deep generative models (variational auto-encoders, GANs, normalizing flows), as well as time-series–specific methods and hybrid/ensemble systems that emphasize robustness, interpretability, and calibrated scoring for real-world monitoring and alerting \citep{chandola2009anomaly,pang2021ano}.

\textit{SKA relevance}: used for transient and FRB discovery in SKA-Low and SKA-Mid data streams (Section~\ref{sec:pulsar_transient_science}), and for identifying rare lensing candidates (Section~\ref{sec:gravitational_lensing}) and unusual source morphologies in continuum surveys (Section~\ref{sec:radio_galaxies}).

\subsubsection{Simulation-based inference (SBI)}\label{sec:sbi}
Simulation-based inference (also called likelihood-free inference) refers to methods for performing statistical inference when the likelihood function is intractable or unavailable but one can forward-simulate data from a generative model; by comparing simulated and observed data via summary statistics, discrepancy measures, or learned neural surrogates, these methods estimate posterior distributions, model parameters, or perform model comparison. Approaches include Approximate Bayesian Computation (ABC), which accepts parameter draws that produce simulations close to observations; synthetic likelihood and density-ratio or conditional density estimation methods that learn likelihood or posterior surrogates; and modern neural approaches (neural posterior estimation, neural likelihood estimation, and neural ratio estimation) that scale simulation-based inference to high-dimensional data using flexible neural networks and amortized inference \citep{beaumont2019approximate,cranmer2020frontier}.

\textit{SKA relevance}: SBI enables parameter inference for EoR cosmology (Section~\ref{sec:eor}), gravitational lensing (Section~\ref{sec:gravitational_lensing}), and pulsar physics (Section~\ref{sec:pulsar_transient_science}) directly from SKA observables without requiring tractable likelihoods -- essential when forward models are complex simulations.


\subsection{Neural network architectures}\label{sec:architectures}

\subsubsection{Convolutional Neural Networks (CNNs)}\label{sec:cnn}
Convolutional neural networks (CNNs) \citep{lecun1998,alexnet} are a class of deep learning models designed to process grid-structured data (especially images) by using convolutional layers that apply learned localized filters across spatial dimensions to extract translation-equivariant features; stacking convolutions, non-linearities, pooling, and normalization produces hierarchical representations that capture edges, textures, parts, and high-level concepts, enabling state-of-the-art performance in image classification, object detection, segmentation, and related tasks.

\textit{SKA relevance}: the most widely used architecture across all six scientific areas reviewed, from RFI detection and source finding to EoR signal extraction.

\subsubsection{Residual Network (ResNet)}\label{sec:resnet}
A Residual Network (ResNet) is a deep convolutional architecture that uses identity-skip (residual) connections to allow training of substantially deeper networks by reformulating layers as learning residual functions \citep{resnet,resnet2}. Each block learns the difference between the desired mapping and the input, thereby mitigating vanishing gradients and enabling easier optimization. ResNets achieved state-of-the-art performance on image classification and became a foundational backbone for many vision tasks and architectures.

\textit{SKA relevance}: used as backbone architecture in many models such as POLISH for interferometric image super-resolution (Section~\ref{sec:radio_interferometric_imaging}) and in 3D magnetic field reconstruction from spectroscopic SKA-Mid observations (Section~\ref{sec:cosmic_magnetism}).

\subsubsection{U-Net}\label{sec:unet}
U-Net is a convolutional encoder–decoder architecture with symmetric contracting and expanding paths connected by skip connections, originally developed for biomedical image segmentation; the encoder progressively captures context via down-sampling convolutions while the decoder restores spatial resolution using up-convolutions, and the lateral skip connections transfer fine-grained localization information from corresponding encoder layers to the decoder, enabling precise pixel-wise predictions from relatively few training images \citep{unet_2015}.

\textit{SKA relevance}: the workhorse architecture for interferometric deconvolution, RFI segmentation (Section~\ref{sec:radio_interferometric_imaging}), EoR foreground separation (Section~\ref{sec:eor}), and many deep generative modelling approaches (Appendix~\ref{sec:generative_models}).

\subsubsection{Transformers}\label{sec:transformer}
Transformers are a neural network architecture that uses self-attention mechanisms to model relationships between all elements of a sequence in parallel, replacing recurrence and enabling efficient scaling to very long contexts \citep{vaswani2017,devlin-etal-2019-bert,gpt3}. By computing attention-weighted combinations of token representations, transformers learn contextualized embeddings that power state-of-the-art results in machine translation, language modelling, and many other domains. For the image modality, Vision Transformers (ViT) \citep{dosovitskiy2020image} have been developed that tokenize images in patches. The model then uses a self-attention mechanism on these patches to capture global relationships across the entire image, providing a powerful alternative to traditional Convolutional Neural Networks (CNNs).

\textit{SKA relevance}: most widely used architecture when data is sufficient, e.g., for uncertainty-aware EoR parameter inference (Section~\ref{sec:eor}), radio source classification and detection (Section~\ref{sec:radio_galaxies}), and gravitational lens modelling (Section~\ref{sec:gravitational_lensing}), where calibrated posteriors are essential for scientific validity.


\subsubsection{Bayesian Neural Network (BNN)}\label{sec:bnn}
Bayesian neural networks (BNNs) \citep{mackay1992,neal2012bayesian} place probability distributions over a network's weights (or outputs) to capture epistemic uncertainty and enable principled Bayesian inference for prediction, calibration, and decision-making; by combining prior beliefs with likelihoods from data, BNNs produce posterior distributions that quantify uncertainty about model parameters and predictions, which is useful for robust learning, active learning, and safety-critical applications. Exact Bayesian inference is intractable for large networks, so practical methods use approximations such as variational inference, Monte Carlo dropout \citep{gal2016dropout}, deep ensembles, and Markov Chain Monte Carlo (MCMC) \citep{Robert2011} to approximate posteriors and predictive distributions.

\textit{SKA relevance}: used for uncertainty-aware EoR parameter inference (Section~\ref{sec:eor}) and gravitational lens modelling (Section~\ref{sec:gravitational_lensing}), where calibrated posteriors are essential for scientific validity.

\subsubsection{Physics-informed neural networks (PINNs)}\label{sec:pinns}
Physics-informed neural networks (PINNs) incorporate known physical laws--typically expressed as differential equations, boundary/initial conditions, and conservation constraints--directly into the training loss so neural networks approximate solutions that satisfy both data and governing equations \citep{raissi2017physicsinformeddeeplearning,raissi2017physicsinformeddeeplearning2,RAISSI2019686,karniadakis2021physics}. By penalizing residuals of the PDE (and related constraints) at collocation points, PINNs enable mesh-free solution of forward and inverse problems, parameter estimation, and surrogate modelling for complex physical systems while often reducing required labelled data and improving generalization.

\textit{SKA relevance}: applied to EoR calibration error quantification (Section~\ref{sec:eor}) and radiative transfer emulation for SKA-Low forward modelling (Section~\ref{sec:radio_interferometric_imaging}).

\subsubsection{You Only Look Once (YOLO)}\label{sec:yolo}
The You Only Look Once (YOLO) family of models revolutionized object detection by formulating it as a single-stage regression problem, enabling simultaneous prediction of bounding boxes and class probabilities with a single pass through the neural network, which drastically increased processing speed \citep{redmon2016you}.

\textit{SKA relevance}: adapted for radio source detection (YOLO-CIANNA) and optical gravitational lens searches with direct applicability to SKA-Mid continuum surveys (Section~\ref{sec:radio_galaxies}).

\subsubsection{Segment Anything (SAM)}\label{sec:sam}
Segment Anything (SAM), a foundational model for computer vision, introduced the concept of promptable segmentation, allowing it to generate high-quality segmentation masks for any object in an image based on simple inputs like clicks, bounding boxes, or text \citep{kirillov2023segment}. Trained on a massive dataset of over one billion masks, SAM exhibits impressive zero-shot generalization capabilities, enabling it to accurately segment objects and novel scenes it has never encountered during training.

\textit{SKA relevance}: applied to automated RFI segmentation in interferometric data for both SKA-Mid and SKA-Low (Section~\ref{sec:radio_interferometric_imaging}).

\subsubsection{Mask R-CNN}\label{sec:mask_rcnn}
Mask R-CNN \citep{he2018maskrcnn} is a deep learning model used for object detection and instance segmentation in images. It extends the Faster R-CNN architecture \citep{ren2016fasterrcnnrealtimeobject} by adding a branch for predicting segmentation masks, allowing it to identify and delineate individual objects within an image accurately.

\textit{SKA relevance}: used for multi-component radio source detection and segmentation in SKA-precursor ASKAP EMU data, with direct applicability to SKA-Mid AA4 continuum surveys (Section~\ref{sec:radio_galaxies}).


\subsection{Deep Generative Models}\label{sec:generative_models}
In recent years, there has been a strong increase in research in the domain of deep learning-based generative models \citep{bondtaylor2022}, in particular for generating high-quality images. Such models showcase impressive generative capabilities, ranging from very high levels of detail to the diversity of the generated samples. The most important types of such models are Generative Adversarial Networks (GANs) \citep{goodfellow2014generative, goodfellow2020generative}, Normalizing Flows (NFs) \citep{dinh2014nice, rezende2015variational, dinh2016density}, Variational Auto-encoders (VAEs) \citep{kingma2013auto}, and Diffusion Models (DMs) \citep{cao2024, yang2023diffusion}. In addition, there are also Auto-regressive \citep{van2016pixel} and Energy-based \citep{lecun2006tutorial, song2021train} models.

\subsubsection{Generative Adversarial Networks (GANs)}\label{sec:gan}
GANs \citep{park2021review, gui2021review} learn to generate data examples $x$ from latent variables $z$, using an adversarial loss that encourages the generated samples to be indistinguishable from real examples in the training data set. GANs consist of two networks, a data generator, and a discriminator which learns to separate generated from real data.

\textit{SKA relevance}: used for sky simulation, map-to-map translation between simulated and observable quantities (SKA-Mid HI science; Section~\ref{sec:radio_interferometric_imaging}), and data augmentation for rare radio morphologies (Section~\ref{sec:radio_galaxies}).

\subsubsection{Normalizing Flows}\label{sec:norm_flow}
Normalizing flows \citep{kobyzev2020normalizing, papamakarios2021normalizing} learn a probability model by transforming a simple distribution into a more complicated one using a deep network. Normalizing flows can both sample from this distribution and evaluate the probability of new examples. However, they require specialized architecture: each layer must be invertible.

\textit{SKA relevance}: used as the density-estimator backbone in simulation-based inference for EoR parameter estimation (Section~\ref{sec:eor}) and gravitational lens modelling (Section~\ref{sec:gravitational_lensing}).

\subsubsection{Variational Auto-encoders (VAEs)}\label{sec:vae}
VAEs \citep{doersch2016tutorial} combine elements of both variational inference and auto-encoding architectures. They learn to encode input data into a latent space distribution and then decode it back to reconstruct the original input, while also optimizing a variational lower bound to ensure the latent space is structured and continuous, allowing for smooth generation of new data points.

\textit{SKA relevance}: used for EoR foreground mitigation via VAE-trained GPR kernels (Section~\ref{sec:eor}), and as anomaly detectors for transient discovery in SKA-Low and SKA-Mid data (Section~\ref{sec:pulsar_transient_science}).

\subsubsection{Diffusion Models (DMs)}\label{sec:diffusion}
Diffusion models are highly flexible probabilistic models that operate in two distinct stages, the forward and reverse diffusion processes. In the forward process, noise is incrementally added to the input, such that at the end of the process the data is transformed into pure Gaussian noise. In the reverse or denoising process, the original input data is gradually recovered from the noise. Diffusion Models fall into three broad categories: Denoising Diffusion Probabilistic Models (DDPM) \citep{sohl2015deep, ho2020denoising}, Score-based Generative Models (SGMs) or Noise Conditioned Score Networks (NCSNs) \citep{song2019generative, song2020improved}, and Stochastic Differential Equations (SDEs) \citep{karras2022elucidating, song2021maximum, song2020score}.

\textit{SKA relevance}: applied across interferometric imaging (Section~\ref{sec:radio_interferometric_imaging}), EoR foreground separation (Section~\ref{sec:eor}), magnetic field estimation (Section~\ref{sec:cosmic_magnetism}), and gravitational lens modelling (Section~\ref{sec:gravitational_lensing}) -- among the most versatile generative architectures for SKA science.

\subsubsection{Comparison of generative models}\label{sec:comparison_generative_models}
Comparing the characteristics of GANs, VAEs, NFs and DMs \citep{xiao2021tackling}, GANs can efficiently generate high quality samples, but their mode coverage is limited. Conversely, NFs and VAEs lack in sampling quality, however with very good mode coverage. To this end, DMs are able to compensate for these deficiencies and offer good mode coverage and high sampling quality at the same time. DMs have shown impressive performance and outperformed GANs in image synthesis \citep{dhariwal2021diffusion}. Nevertheless, due to their iterative nature, which causes a slow sampling process, they are computationally expensive.

\subsubsection{Conditional Generation}\label{sec:cond_generation}
Generative models can generate new samples such as images in an unconditional setting, while being trained in an unsupervised way. Perhaps more interesting for scientific applications is conditional image generation. The conditioning can be based on attributes such as class labels, text prompts or other images. The image-conditioned setting leads to the task of image-to-image (I2I) translation. Here, the model is conditioned on an input image y, such that for a probabilistic generative model the probability for each data point $p(x)$ becomes $p(x|y)$. The model can learn a transformation from a noisy, low-resolution, or otherwise degraded input image to a higher quality output image, often using a training data set of paired images. In computer vision, this could be denoising, super-resolution, style transfer, recolouring, etc.

\textit{SKA relevance}: the image-to-image translation paradigm underlies map-to-map translation between simulation and observational domains, essential for bridging hydrodynamical simulations with SKA-Mid observations (Section~\ref{sec:radio_interferometric_imaging}).

\subsubsection{Pix2pix and CycleGAN}\label{sec:cgan}
In \citep{isola2017image} (Pix2pix), the problem of I2I translation was tackled using a conditional GAN derived from the DCGAN \citep{radford2015unsupervised} architecture: A U-Net \citep{unet_2015} architecture is used as a generator, which receives the noise input concatenated with the conditioning image, and a patch discriminator. In \citep{zhu2017unpaired} (CycleGAN), this approach was extended to unpaired training data. It uses two mapping functions $G:X\rightarrow Y$ and $F:Y\rightarrow X$ with associated adversarial discriminators, as well as a cycle consistency loss to enforce $F(G(X)) \sim X$ (and vice versa).

\textit{SKA relevance}: conditional GAN frameworks including Pix2pix and CycleGAN have been applied to radio galaxy image synthesis and map-to-map translation tasks relevant for SKA-Mid forward modelling (Section~\ref{sec:radio_interferometric_imaging} and \ref{sec:eor}).

\subsubsection{Flow Matching models}\label{sec:flow_matching}
Flow matching models \citep{albergo2023buildingnormalizingflowsstochastic,albergo2023stochasticinterpolantsunifyingframework,liu2022flowstraightfastlearning,lipman2023flowmatchinggenerativemodeling} represent an innovative approach in generative modelling, merging the strengths of Continuous Normalizing Flows (CNFs) and diffusion models to address key limitations of both methodologies. 
Unlike traditional flow-based models that explicitly parametrize invertible mappings, flow matching focuses on directly estimating the velocity fields that transport one distribution to another, often by solving an ordinary differential equation (ODE) that characterizes the flow dynamics. This approach enables efficient and flexible generation by matching the probability flow paths without requiring explicit invertibility, thereby reducing computational complexity and training difficulty. Recent developments have extended flow matching to various modalities, including images and audio \citep{lipman2024flowmatchingguidecode,holderrieth2025introductionflowmatchingdiffusion}.

\textit{SKA relevance}: flow matching architectures are applicable to pulse profile denoising for SKA pulsar timing arrays (Section~\ref{sec:pulsar_transient_science}), and as computationally efficient alternatives to diffusion models for sky simulation and image reconstruction tasks (Section~\ref{sec:radio_interferometric_imaging}).

\newcommand{\actaa}{Acta Astron.} 
\newcommand{\araa}{ARA\&A} 
\newcommand{\aar}{A\&ARv} 
\newcommand{\aapr}{A\&ARv} 
\newcommand{\ab}{Astrobiol.} 
\newcommand{\aj}{AJ} 
\newcommand{\apj}{ApJ} 
\newcommand{\apjl}{ApJL} 
\newcommand{\apjs}{ApJSS} 
\newcommand{\ao}{Appl. Opt.} 
\newcommand{\apss}{Astro. \& Space Sci.} 
\newcommand{\aap}{A\&A} 
\newcommand{\aaps}{A\&AS.} 
\newcommand{\baas}{Bull. Am. Astron. Soc.} 
\newcommand{\caa}{Chinese A\&A} 
\newcommand{\cjaa}{Chinese J. A\&A} 
\newcommand{\cqg}{Class. Quantum Gravity} 
\newcommand{\gal}{Galaxies} 
\newcommand{\gca}{Geo. Cosmo. Acta} 
\newcommand{\icarus}{Icarus} 
\newcommand{\jcap}{JCAP} 
\newcommand{\jgr}{J. Geophys. Res.} 
\newcommand{\jgrp}{J. Geophys. Res. Planets} 
\newcommand{\jqsrt}{J. Quant. Spectrosc. Radiat. Transf.} 
\newcommand{\memsai}{Mem. SAIt} 
\newcommand{\mnras}{MNRAS} 
\newcommand{\nat}{Nature} 
\newcommand{\nastro}{Nat. Astron.} 
\newcommand{\ncomms}{Nat. Commun.} 
\newcommand{\nphys}{Nat. Phys.} 
\newcommand{\na}{New Astron.} 
\newcommand{\nar}{New Astron. Rev.} 
\newcommand{\physrep}{Phys. Rep.} 
\newcommand{\pra}{Phys. Rev. A} 
\newcommand{\prb}{Phys. Rev. B} 
\newcommand{\prc}{Phys. Rev. C} 
\newcommand{\prd}{Phys. Rev. D} 
\newcommand{\pre}{Phys. Rev. E} 
\newcommand{\prx}{Phys. Rev. X} 
\newcommand{\prl}{Phys. Rev. Let.} 
\newcommand{\psj}{Planet. Sci. J.} 
\newcommand{\planss}{Planet. Space Sci.} 
\newcommand{\pnas}{Proc. Natl Acad. Sci. USA} 
\newcommand{\procspie}{Proc. SPIE} 
\newcommand{\pasa}{PASA} 
\newcommand{\pasj}{PASJ} 
\newcommand{\pasp}{PASP} 
\newcommand{\rmxaa}{RMXAA} 
\newcommand{\sci}{Science} 
\newcommand{\sciadv}{Sci. Adv.} 
\newcommand{\solphys}{Sol. Phys.} 
\newcommand{\sovast}{Soviet Ast.} 
\newcommand{\ssr}{Space Sci. Rev.} 
\newcommand{\uni}{Universe} 

\bibliographystyle{abbrvnat-maxbibnames4}
\bibliography{chapter, SKABook_gava, SKABook_denp, aaska2_crossref} 

\begin{thebibliography}{225}
\providecommand{\natexlab}[1]{#1}
\providecommand{\url}[1]{\texttt{#1}}
\expandafter\ifx\csname urlstyle\endcsname\relax
  \providecommand{\doi}[1]{doi: #1}\else
  \providecommand{\doi}{doi: \begingroup \urlstyle{rm}\Url}\fi

\bibitem[Acharya et~al.(2024{\natexlab{a}})Acharya, Mertens, Ciardi, Ghara,
  Koopmans, Giri, Hothi, Ma, Mellema, and Munshi]{acharya21cmSignalEpoch2024}
A.~Acharya et al.
\newblock \emph{Mon Not R Astron Soc}, 527\penalty0 (3):\penalty0 7835--7846,
  Jan. 2024{\natexlab{a}}.
\newblock ISSN 0035-8711.
\newblock \doi{10.1093/mnras/stad3701}.
\newblock URL \url{https://doi.org/10.1093/mnras/stad3701}.

\bibitem[Acharya et~al.(2024{\natexlab{b}})Acharya, Mertens, Ciardi, Ghara,
  Koopmans, and Zaroubi]{acharyaRevisedLOFARUpper2024}
A.~Acharya et al.
\newblock \emph{Mon Not R Astron Soc Lett}, 534\penalty0 (1):\penalty0
  L30--L34, Oct. 2024{\natexlab{b}}.
\newblock ISSN 1745-3925.
\newblock \doi{10.1093/mnrasl/slae078}.
\newblock URL \url{https://doi.org/10.1093/mnrasl/slae078}.

\bibitem[Acharya et~al.(2026{\natexlab{a}})Acharya, author2, author3, author4,
  and author5]{Acharya01.2026.SKA}
A.~Acharya et al.
\newblock In \emph{Advancing Astrophysics with the SKA -- II (AASKAII)}.
  2026{\natexlab{a}}.
\newblock arXiv search: Report number AASKAII/Acharya01.

\bibitem[Acharya et~al.(2026{\natexlab{b}})Acharya, author2, author3, author4,
  and author5]{Acharya02.2026.SKA}
A.~Acharya et al.
\newblock In \emph{Advancing Astrophysics with the SKA -- II (AASKAII)}.
  2026{\natexlab{b}}.
\newblock arXiv search: Report number AASKAII/Acharya02.

\bibitem[Agarwal et~al.(2020)Agarwal, Aggarwal, Burke-Spolaor, Lorimer, and
  Garver-Daniels]{agarwal_2020_fetch}
D.~Agarwal et al.
\newblock \emph{Monthly Notices of the Royal Astronomical Society},
  497\penalty0 (2):\penalty0 1661--1674, 09 2020.
\newblock ISSN 0035-8711.
\newblock \doi{10.1093/mnras/staa1856}.
\newblock URL \url{https://doi.org/10.1093/mnras/staa1856}.

\bibitem[Aghabiglou et~al.(2024)Aghabiglou, Chu, Dabbech, and
  Wiaux]{aghabiglou_2024_r2d2_deep}
A.~Aghabiglou, C.~S. Chu, A.~Dabbech, and Y.~Wiaux.
\newblock \emph{The Astrophysical Journal Supplement Series}, 273\penalty0
  (1):\penalty0 3, 2024.
\newblock \doi{10.3847/1538-4365/ad46f5}.
\newblock URL \url{http://dx.doi.org/10.3847/1538-4365/ad46f5}.

\bibitem[Aghabiglou et~al.(2025)Aghabiglou, Chu, Tang, Dabbech, and
  Wiaux]{aghabiglou_2025_toward_robust}
A.~Aghabiglou et al.
\newblock \emph{The Astrophysical Journal Supplement Series}, 280\penalty0
  (2):\penalty0 63, 2025.
\newblock \doi{10.3847/1538-4365/adfbed}.
\newblock URL \url{http://dx.doi.org/10.3847/1538-4365/adfbed}.

\bibitem[Akeret et~al.(2017)Akeret, Chang, Lucchi, and
  Refregier]{akeret_2017_radio_frequency}
J.~Akeret, C.~Chang, A.~Lucchi, and A.~Refregier.
\newblock \emph{Astronomy and Computing}, 18:\penalty0 35--39, 2017.
\newblock \doi{10.1016/j.ascom.2017.01.002}.
\newblock URL \url{http://dx.doi.org/10.1016/j.ascom.2017.01.002}.

\bibitem[Albergo and
  Vanden-Eijnden(2023)]{albergo2023buildingnormalizingflowsstochastic}
M.~S. Albergo and E.~Vanden-Eijnden.
\newblock Building normalizing flows with stochastic interpolants, 2023.
\newblock URL \url{https://arxiv.org/abs/2209.15571}.

\bibitem[Albergo et~al.(2023)Albergo, Boffi, and
  Vanden-Eijnden]{albergo2023stochasticinterpolantsunifyingframework}
M.~S. Albergo, N.~M. Boffi, and E.~Vanden-Eijnden.
\newblock Stochastic interpolants: A unifying framework for flows and
  diffusions, 2023.
\newblock URL \url{https://arxiv.org/abs/2303.08797}.

\bibitem[Anderson et~al.(2026)Anderson, author2, author3, author4, and
  author5]{GemmaAnderson01.2026.SKA}
G.~E. Anderson et al.
\newblock In \emph{Advancing Astrophysics with the SKA -- II (AASKAII)}. 2026.
\newblock arXiv search: Report number AASKAII/GemmaAnderson01.

\bibitem[Andersson et~al.(2025)Andersson, Lintott, Fender, Lochner, Woudt,
  {van~den~Eijnden}, {van~der~Horst}, Horesh, Saikia, Sivakoff, Tremou, and
  Vaccari]{anderssonFindingRadioTransients2025}
A.~Andersson et al.
\newblock \emph{Mon Not R Astron Soc}, 538\penalty0 (3):\penalty0 1397--1414,
  Apr. 2025.
\newblock ISSN 0035-8711.
\newblock \doi{10.1093/mnras/staf336}.
\newblock URL \url{https://doi.org/10.1093/mnras/staf336}.

\bibitem[Andersson et~al.(2026{\natexlab{a}})Andersson, author2, author3,
  author4, and author5]{AlexAndersson01.2026.SKA}
A.~Andersson et al.
\newblock In \emph{Advancing Astrophysics with the SKA -- II (AASKAII)}.
  2026{\natexlab{a}}.
\newblock arXiv search: Report number AASKAII/AlexAndersson01.

\bibitem[Andersson et~al.(2026{\natexlab{b}})Andersson, author2, author3,
  author4, and author5]{AlexAndersson02.2026.SKA}
A.~Andersson et al.
\newblock In \emph{Advancing Astrophysics with the SKA -- II (AASKAII)}.
  2026{\natexlab{b}}.
\newblock arXiv search: Report number AASKAII/AlexAndersson02.

\bibitem[{Baron Perez} et~al.(2025){Baron Perez}, {Br\"uggen}, {Kasieczka}, and
  {Lucie-Smith}]{baron_perez_2025_classification_radio}
N.~{Baron Perez}, M.~{Br\"uggen}, G.~{Kasieczka}, and L.~{Lucie-Smith}.
\newblock \emph{A\&A}, 699:\penalty0 A302, 2025.
\newblock \doi{10.1051/0004-6361/202554735}.
\newblock URL \url{https://doi.org/10.1051/0004-6361/202554735}.

\bibitem[Beaumont(2019)]{beaumont2019approximate}
M.~A. Beaumont.
\newblock \emph{Annual review of statistics and its application}, 6\penalty0
  (1):\penalty0 379--403, 2019.
\newblock \doi{10.1146/annurev-statistics-030718-105212}.
\newblock URL \url{https://doi.org/10.1146/annurev-statistics-030718-105212}.

\bibitem[Bernardini et~al.(2021)Bernardini, Feldmann, Angl{\'e}s-Alc{\'a}zar,
  Boylan-Kolchin, Bullock, Mayer, and Stadel]{bernardini_2021_from_ember}
M.~Bernardini et al.
\newblock \emph{\mnras}, 509\penalty0 (1):\penalty0 1323--1341, 2021.
\newblock \doi{10.1093/mnras/stab3088}.
\newblock URL \url{http://dx.doi.org/10.1093/mnras/stab3088}.

\bibitem[Bernardini et~al.(2025)Bernardini, Feldmann, Gensior,
  Anglés-Alcázar, Bassini, Bieri, Cenci, Tortora, and
  Faucher-Giguère]{bernardini_2025_ember_2}
M.~Bernardini et al.
\newblock \emph{Monthly Notices of the Royal Astronomical Society},
  538\penalty0 (2):\penalty0 1201--1215, 02 2025.
\newblock ISSN 0035-8711.
\newblock \doi{10.1093/mnras/staf341}.
\newblock URL \url{https://doi.org/10.1093/mnras/staf341}.

\bibitem[Bernardini et~al.(2026)Bernardini, Feldmann, Anglés-Alcázar, Denzel,
  and Gensior]{bernardini_2026_ember_hi}
M.~Bernardini et al.
\newblock \emph{Monthly Notices of the Royal Astronomical Society}, page
  stag204, 03 2026.
\newblock ISSN 0035-8711.
\newblock \doi{10.1093/mnras/stag204}.
\newblock URL \url{https://doi.org/10.1093/mnras/stag204}.

\bibitem[Bianco et~al.(2025)Bianco, Giri, Sharma, Chen, Krishna, Finlay,
  Nistane, Denzel, Santis, and Ghorbel]{bianco_2025_deep_learning}
M.~Bianco et al.
\newblock \emph{Monthly Notices of the Royal Astronomical Society},
  541\penalty0 (1):\penalty0 234--250, 2025.
\newblock \doi{10.1093/mnras/staf973}.
\newblock URL \url{http://dx.doi.org/10.1093/mnras/staf973}.

\bibitem[Billeter et~al.(2024)Billeter, Denzel, Chavarriaga, Forster,
  Schilling, Brunner, Frischknecht-Gruber, Reif, and
  Weng]{billeter_2024_mlops_enabler}
Y.~Billeter et al.
\newblock In \emph{2024 11th IEEE Swiss Conference on Data Science (SDS)},
  pages 37--40, 5 2024.
\newblock \doi{10.1109/sds60720.2024.00013}.
\newblock URL \url{http://dx.doi.org/10.1109/SDS60720.2024.00013}.

\bibitem[Bonaldi and Braun(2018)]{bonaldi_2018_square_kilometre}
A.~Bonaldi and R.~Braun.
\newblock \emph{CoRR}, 2018.
\newblock URL \url{http://arxiv.org/abs/1811.10454v1}.

\bibitem[Bonaldi et~al.(2025)Bonaldi, Hartley, Braun, Purser, Acharya, Ahn,
  Resco, Bait, Bianco, Chakraborty, Chapman, Chatterjee, Chege, Chen, Chen,
  Chen, Conaboy, Cruz, Darriba, Santis, Denzel, Diao, Feron, Finlay, Gehlot,
  Ghosh, Giri, Grumitt, Hong, Ito, Jiang, Jordan, Kim, Kim, Kim, Krishna,
  Kulkarni, L{\'o}pez-Caniego, Labadie-Garc{\'i}a, Lee, Lee, Lee, Line, Liu,
  Mao, Mazumder, Mertens, Munshi, Nasirudin, Ni, Nistane, Norregaard, Null,
  Offringa, Oh, Oh, Parkinson, Pritchard, Ruiz-Granda, L{\'o}pez, Shan, Sharma,
  Trott, Yoshiura, Zhang, Zhang, Zheng, Zhu, Zuo, Akahori, Alberto, Allys, An,
  Anstey, Baek, Basavraj, Brackenhoff, Browne, Ceccotti, Chen, Chen, Choudhuri,
  Choudhury, Coles, Cook, Cornu, Cunnington, Das, de~Lera Acedo~Acedo, Delouis,
  Deng, Ding, Elahi, Fernandez, Fern{\'a}ndez, Alc{\'a}zar, Galluzzi, Gao,
  Garain, Garrido, Gendron-Marsolais, Gessey-Jones, Ghorbel, Gong, Guo,
  Hasegawa, Hayashi, Herranz, Holanda, Holloway, Hothi, H{\"o}fer, Jeli{\'c},
  Jiang, Jiang, Kang, Kim, Koopmans, Lacroix, Lee, Leeney, Levrier, Li, Liu,
  Ma, Meriot, Mesinger, Mevius, Minoda, Miville-Deschênes, Moldon, Mondal,
  Murmu, Murray, SR, Niu, Nunhokee, O'Hara, Pal, Pal, Park, Parra, Patra,
  Pindor, Remazeilles, Rey, Rubino-Martin, Saha, Selvaraj, Semelin, Shah, Shao,
  Shaw, Shi, Shimabukuro, Singh, Sohn, Stagni, Starck, Sui, Swinbank,
  S{\'a}nchez, S{\'a}nchez-Exp{\'o}sito, Takahashi, Takeuchi, Tripathi,
  Verdes-Montenegro, Vielva, Vitello, Wang, Wang, Wang, Wang, Wang, Wiegert,
  Wild, Williams, Wolz, Wu, Wu, Xia, Xu, Yan, Yan, Yin, You, Yu, Yu, Yue,
  Zhang, Zhao, Zhao, and Zhou]{bonaldi_2025_square_kilometre}
A.~Bonaldi et al.
\newblock \emph{Monthly Notices of the Royal Astronomical Society},
  543\penalty0 (2):\penalty0 1092--1119, 2025.
\newblock \doi{10.1093/mnras/staf1466}.
\newblock URL \url{http://dx.doi.org/10.1093/mnras/staf1466}.

\bibitem[Bond-Taylor et~al.(2022)Bond-Taylor, Leach, Long, and
  Willcocks]{bondtaylor2022}
S.~Bond-Taylor, A.~Leach, Y.~Long, and C.~G. Willcocks.
\newblock \emph{IEEE Transactions on Pattern Analysis and Machine
  Intelligence}, 44\penalty0 (11):\penalty0 7327--7347, 2022.
\newblock \doi{10.1109/TPAMI.2021.3116668}.
\newblock URL \url{https://doi.org/10.1109/TPAMI.2021.3116668}.

\bibitem[Bourke et~al.(2026)Bourke, author2, author3, author4, and
  author5]{Bourke01.2026.SKA}
T.~L. Bourke et al.
\newblock In \emph{Advancing Astrophysics with the SKA -- II (AASKAII)}. 2026.
\newblock arXiv search: Report number AASKAII/Bourke01.

\bibitem[Breck et~al.(2017)Breck, Cai, Nielsen, Salib, and Sculley]{Breck2017}
E.~Breck et al.
\newblock In \emph{2017 IEEE International Conference on Big Data (Big Data)},
  pages 1123--1132. IEEE, Dec. 2017.
\newblock \doi{10.1109/bigdata.2017.8258038}.
\newblock URL \url{https://api.semanticscholar.org/CorpusID:6244440}.

\bibitem[Breitman et~al.(2023)Breitman, Mesinger, Murray, Prelogovi{\'c}, Qin,
  and Trotta]{breitman_2023_21cm_emu}
D.~Breitman et al.
\newblock \emph{Monthly Notices of the Royal Astronomical Society},
  527\penalty0 (4):\penalty0 9833--9852, 2023.
\newblock \doi{10.1093/mnras/stad3849}.
\newblock URL \url{http://dx.doi.org/10.1093/mnras/stad3849}.

\bibitem[Brown et~al.(2020)Brown, Mann, Ryder, Subbiah, Kaplan, Dhariwal,
  Neelakantan, Shyam, Sastry, Askell, Agarwal, Herbert-Voss, Krueger, Henighan,
  Child, Ramesh, Ziegler, Wu, Winter, Hesse, Chen, Sigler, Litwin, Gray, Chess,
  Clark, Berner, McCandlish, Radford, Sutskever, and Amodei]{gpt3}
T.~B. Brown et al.
\newblock In \emph{Proceedings of the 34th International Conference on Neural
  Information Processing Systems}, NIPS '20, Red Hook, NY, USA, 2020. Curran
  Associates Inc.
\newblock ISBN 9781713829546.
\newblock URL
  \url{https://proceedings.neurips.cc/paper/2020/file/1457c0d6bfcb4967418bfb8ac142f64a-Paper.pdf}.

\bibitem[Burba et~al.(2026)Burba, author2, author3, author4, and
  author5]{Burba01.2026.SKA}
J.~Burba et al.
\newblock In \emph{Advancing Astrophysics with the SKA -- II (AASKAII)}. 2026.
\newblock arXiv search: Report number AASKAII/Burba01.

\bibitem[Burgay et~al.(2026)Burgay, Karastergiou, author3, author4, and
  author5]{Joshi01.2026.SKA}
M.~Burgay et al.
\newblock In \emph{Advancing Astrophysics with the SKA -- II (AASKAII)}. 2026.
\newblock arXiv search: Report number AASKAII/Joshi01.

\bibitem[Burke et~al.(2019)Burke, Aleo, Chen, Liu, Peterson, Sembroski, and
  Lin]{burke_2019_deblending_classifying}
C.~J. Burke et al.
\newblock \emph{Monthly Notices of the Royal Astronomical Society},
  490\penalty0 (3):\penalty0 3952--3965, 2019.
\newblock \doi{10.1093/mnras/stz2845}.
\newblock URL \url{http://dx.doi.org/10.1093/mnras/stz2845}.

\bibitem[Cao et~al.(2024)Cao, Tan, Gao, Xu, Chen, Heng, and Li]{cao2024}
H.~Cao et al.
\newblock \emph{IEEE Transactions on Knowledge and Data Engineering}, pages
  1--20, 2024.
\newblock \doi{10.1109/TKDE.2024.3361474}.
\newblock URL \url{https://doi.org/10.1109/TKDE.2024.3361474}.

\bibitem[C{\'a}rcamo et~al.(2022)C{\'a}rcamo, Scaife, Alexander, and
  Leahy]{carcamo_2022_cs_romer}
M.~C{\'a}rcamo, A.~M.~M. Scaife, E.~L. Alexander, and J.~P. Leahy.
\newblock \emph{Monthly Notices of the Royal Astronomical Society},
  518\penalty0 (2):\penalty0 1955--1974, 2022.
\newblock \doi{10.1093/mnras/stac3031}.
\newblock URL \url{http://dx.doi.org/10.1093/mnras/stac3031}.

\bibitem[Carcamo et~al.(2026)Carcamo, author2, author3, author4, and
  author5]{Carcamo01.2026.SKA}
M.~Carcamo et al.
\newblock In \emph{Advancing Astrophysics with the SKA -- II (AASKAII)}. 2026.
\newblock arXiv search: Report number AASKAII/Carcamo01.

\bibitem[Caron et~al.(2021)Caron, Touvron, Misra, Jégou, Mairal, Bojanowski,
  and Joulin]{caron2021emergingpropertiesselfsupervisedvision}
M.~Caron et al.
\newblock Emerging properties in self-supervised vision transformers, 2021.
\newblock URL \url{https://arxiv.org/abs/2104.14294}.

\bibitem[Carretti and Vazza(2025)]{carretti_2025_radio_observations}
E.~Carretti and F.~Vazza.
\newblock \emph{Universe}, 11\penalty0 (5):\penalty0 164, 2025.
\newblock \doi{10.3390/universe11050164}.
\newblock URL \url{http://dx.doi.org/10.3390/universe11050164}.

\bibitem[Carrillo et~al.(2012)Carrillo, McEwen, and
  Wiaux]{carrillo_2012_sparsity_averaging}
R.~E. Carrillo, J.~D. McEwen, and Y.~Wiaux.
\newblock \emph{Monthly Notices of the Royal Astronomical Society},
  426\penalty0 (2):\penalty0 1223--1234, 2012.
\newblock \doi{10.1111/j.1365-2966.2012.21605.x}.
\newblock URL \url{http://dx.doi.org/10.1111/j.1365-2966.2012.21605.x}.

\bibitem[Ceccotti et~al.(2025)Ceccotti, Offringa, Mertens, Koopmans, Munshi,
  Chege, Acharya, Brackenhoff, Chapman, Ciardi, Ghara, Ghosh, Giri, H{\"o}fer,
  Hothi, Mellema, Mevius, Pandey, and Zaroubi]{ceccottiFirstUpperLimits2025b}
E.~Ceccotti et al.
\newblock \emph{Mon Not R Astron Soc}, page staf1629, Sept. 2025.
\newblock ISSN 0035-8711.
\newblock \doi{10.1093/mnras/staf1629}.
\newblock URL \url{https://doi.org/10.1093/mnras/staf1629}.

\bibitem[Cerardi et~al.(2025)Cerardi, Giri, Bianco, Piras, Salis, Santis,
  Selcuk-Simsek, Denzel, Hess, Toribio, Kirsten, and
  Ghorbel]{cerardi_2025_implicit_inference}
N.~Cerardi et al.
\newblock \emph{CoRR}, 2025.
\newblock URL \url{http://arxiv.org/abs/2511.11568v1}.

\bibitem[Chandola et~al.(2009)Chandola, Banerjee, and
  Kumar]{chandola2009anomaly}
V.~Chandola, A.~Banerjee, and V.~Kumar.
\newblock \emph{ACM computing surveys (CSUR)}, 41\penalty0 (3):\penalty0 1--58,
  2009.
\newblock \doi{10.1145/1541880.1541882}.
\newblock URL \url{https://doi.org/10.1145/1541880.1541882}.

\bibitem[Cheng et~al.(2020)Cheng, Li, Conselice, Arag{\'o}n-Salamanca, Dye, and
  Metcalf]{cheng_2020_identifying_strong}
T.-Y. Cheng et al.
\newblock \emph{Monthly Notices of the Royal Astronomical Society},
  494\penalty0 (3):\penalty0 3750--3765, 2020.
\newblock \doi{10.1093/mnras/staa1015}.
\newblock URL \url{http://dx.doi.org/10.1093/mnras/staa1015}.

\bibitem[Choudhury et~al.(2022)Choudhury, Datta, and
  Majumdar]{choudhuryExtracting21cmPower2022}
M.~Choudhury, A.~Datta, and S.~Majumdar.
\newblock \emph{Mon Not R Astron Soc}, 512\penalty0 (4):\penalty0 5010--5022,
  June 2022.
\newblock ISSN 0035-8711.
\newblock \doi{10.1093/mnras/stac736}.
\newblock URL \url{https://doi.org/10.1093/mnras/stac736}.

\bibitem[Connor and van Leeuwen(2018)]{connor_2018_applying_deep}
L.~Connor and J.~van Leeuwen.
\newblock \emph{The Astronomical Journal}, 156\penalty0 (6):\penalty0 256,
  2018.
\newblock \doi{10.3847/1538-3881/aae649}.
\newblock URL \url{http://dx.doi.org/10.3847/1538-3881/aae649}.

\bibitem[Connor et~al.(2022)Connor, Bouman, Ravi, and
  Hallinan]{connor_2022_deep_radio}
L.~Connor, K.~L. Bouman, V.~Ravi, and G.~Hallinan.
\newblock \emph{Monthly Notices of the Royal Astronomical Society},
  514\penalty0 (2):\penalty0 2614--2626, 2022.
\newblock \doi{10.1093/mnras/stac1329}.
\newblock URL \url{http://dx.doi.org/10.1093/mnras/stac1329}.

\bibitem[Cornu et~al.(2024)Cornu, Salom{\'e}, Semelin, Marchal, Freundlich,
  Aicardi, Lu, Sainton, Mertens, Combes, and Tasse]{cornu_2024_yolo_cianna}
D.~Cornu et al.
\newblock \emph{Astronomy \& Astrophysics}, 690:\penalty0 A211, 2024.
\newblock \doi{10.1051/0004-6361/202449548}.
\newblock URL \url{http://dx.doi.org/10.1051/0004-6361/202449548}.

\bibitem[{Cornu} et~al.(2026){Cornu}, {Semelin}, {Salom\'e}, {Lu}, {Aicardi},
  {Freundlich}, {Mertens}, {Marchal}, {Sainton}, {Combes}, and
  {Tasse}]{cornu_2026_yolo_cianna}
D.~{Cornu} et al.
\newblock \emph{A\&A}, 707:\penalty0 A203, 2026.
\newblock \doi{10.1051/0004-6361/202557257}.
\newblock URL \url{https://doi.org/10.1051/0004-6361/202557257}.

\bibitem[Cranmer et~al.(2020)Cranmer, Brehmer, and Louppe]{cranmer2020frontier}
K.~Cranmer, J.~Brehmer, and G.~Louppe.
\newblock \emph{Proceedings of the National Academy of Sciences}, 117\penalty0
  (48):\penalty0 30055--30062, 2020.
\newblock \doi{10.1073/pnas.1912789117}.
\newblock URL \url{https://doi.org/10.1073/pnas.1912789117}.

\bibitem[de~Salis et~al.(2025)de~Salis, Santis, Piras, Giri, Bianco, Cerardi,
  Denzel, Selcuk-Simsek, Hess, Toribio, Kirsten, and
  Ghorbel]{salis_2025_exploring_early}
E.~de~Salis et al.
\newblock \emph{Exploring the Early Universe with Deep Learning}, pages
  426--438.
\newblock Lecture Notes in Computer Science. Springer Nature Switzerland, 2025.
\newblock \doi{10.1007/978-3-032-05176-9_33}.
\newblock URL \url{http://dx.doi.org/10.1007/978-3-032-05176-9_33}.

\bibitem[Deal and Jagannathan(2024)]{deal_2024_segmenting_rfi}
D.~Deal and P.~Jagannathan.
\newblock \emph{Research Notes of the AAS}, 8\penalty0 (10):\penalty0 257,
  2024.
\newblock \doi{10.3847/2515-5172/ad84fa}.
\newblock URL \url{http://dx.doi.org/10.3847/2515-5172/ad84fa}.

\bibitem[Deng et~al.(2009)Deng, Dong, Socher, Li, Li, and
  Fei-Fei]{deng2009imagenet}
J.~Deng et al.
\newblock In \emph{2009 IEEE conference on computer vision and pattern
  recognition}, pages 248--255. Ieee, 2009.
\newblock \doi{10.1109/cvpr.2009.5206848}.
\newblock URL \url{https://doi.org/10.1109/cvpr.2009.5206848}.

\bibitem[Denzel et~al.(2024)Denzel, Brunner, Billeter, Forster,
  Frischknecht-Gruber, Reif, Schilling, Weng, Chavarriaga, Amini, Repetto, and
  Iranfar]{denzel_2024_towards_certification}
P.~Denzel et al.
\newblock In \emph{2024 11th IEEE Swiss Conference on Data Science (SDS)},
  pages 84--91, 2024.
\newblock \doi{10.1109/SDS60720.2024.00020}.
\newblock URL \url{https://doi.org/10.1109/SDS60720.2024.00020}.

\bibitem[Denzel et~al.(2026)Denzel, Billeter, Schilling, and
  Gavagnin]{denzel_2025_galactic_alchemy}
P.~Denzel, Y.~Billeter, F.-P. Schilling, and E.~Gavagnin.
\newblock \emph{Monthly Notices of the Royal Astronomical Society},
  546\penalty0 (4):\penalty0 stag155, 02 2026.
\newblock ISSN 0035-8711.
\newblock \doi{10.1093/mnras/stag155}.
\newblock URL \url{https://doi.org/10.1093/mnras/stag155}.

\bibitem[Devlin et~al.(2019)Devlin, Chang, Lee, and
  Toutanova]{devlin-etal-2019-bert}
J.~Devlin, M.-W. Chang, K.~Lee, and K.~Toutanova.
\newblock In J.~Burstein, C.~Doran, and T.~Solorio, editors, \emph{Proceedings
  of the 2019 Conference of the North {A}merican Chapter of the Association for
  Computational Linguistics: Human Language Technologies, Volume 1 (Long and
  Short Papers)}, pages 4171--4186, Minneapolis, Minnesota, June 2019.
  Association for Computational Linguistics.
\newblock \doi{10.18653/v1/N19-1423}.
\newblock URL \url{https://aclanthology.org/N19-1423/}.

\bibitem[Dhariwal and Nichol(2021)]{dhariwal2021diffusion}
P.~Dhariwal and A.~Nichol.
\newblock In \emph{Proceedings of the 35th International Conference on Neural
  Information Processing Systems}, NIPS '21, Red Hook, NY, USA, 2021. Curran
  Associates Inc.
\newblock ISBN 9781713845393.
\newblock URL
  \url{https://proceedings.neurips.cc/paper/2021/file/49ad23d1ec9fa4bd8d77d02681df5cfa-Paper.pdf}.

\bibitem[Dinh et~al.(2014)Dinh, Krueger, and Bengio]{dinh2014nice}
L.~Dinh, D.~Krueger, and Y.~Bengio.
\newblock \emph{arXiv preprint arXiv:1410.8516}, 2014.
\newblock \doi{10.48550/arXiv.1410.8516}.
\newblock URL \url{https://doi.org/10.48550/arXiv.1410.8516}.

\bibitem[Dinh et~al.(2016)Dinh, Sohl-Dickstein, and Bengio]{dinh2016density}
L.~Dinh, J.~Sohl-Dickstein, and S.~Bengio.
\newblock \emph{arXiv preprint arXiv:1605.08803}, 2016.
\newblock \doi{10.48550/arXiv.1605.08803}.
\newblock URL \url{https://doi.org/10.48550/arXiv.1605.08803}.

\bibitem[Doersch(2016)]{doersch2016tutorial}
C.~Doersch.
\newblock \emph{arXiv preprint arXiv:1606.05908}, 2016.
\newblock \doi{10.48550/arXiv.1606.05908}.
\newblock URL \url{https://doi.org/10.48550/arXiv.1606.05908}.

\bibitem[Dosovitskiy et~al.(2020)Dosovitskiy, Beyer, Kolesnikov, Weissenborn,
  Zhai, Unterthiner, Dehghani, Minderer, Heigold, Gelly, Uszkoreit, and
  Houlsby]{dosovitskiy2020image}
A.~Dosovitskiy et al.
\newblock \emph{arXiv preprint arXiv:2010.11929}, 2020.
\newblock \doi{10.48550/arXiv.2010.11929}.
\newblock URL \url{https://doi.org/10.48550/arXiv.2010.11929}.

\bibitem[Drozdova et~al.(2024)Drozdova, Kinakh, Bait, Taran, Lastufka,
  Dessauges-Zavadsky, Holotyak, Schaerer, and
  Voloshynovskiy]{drozdova_2024_radio_astronomical}
M.~Drozdova et al.
\newblock \emph{Astronomy \& Astrophysics}, 683:\penalty0 A105, 2024.
\newblock \doi{10.1051/0004-6361/202347948}.
\newblock URL \url{http://dx.doi.org/10.1051/0004-6361/202347948}.

\bibitem[Drozdova et~al.(2025)Drozdova, Lastufka, Kinakh, Holotyak, Schaerer,
  and Voloshynovskiy]{drozdova_2025_radio_astronomy}
M.~Drozdova et al.
\newblock \emph{CoRR}, 2025.
\newblock URL \url{http://arxiv.org/abs/2509.02615v2}.

\bibitem[Dulwich and Mort(2017)]{dulwich_2017_oskar}
F.~Dulwich and B.~Mort.
\newblock \emph{OSKAR 2.7.0}.
\newblock Zenodo, 2017.
\newblock \doi{10.5281/ZENODO.3758449}.
\newblock URL \url{https://zenodo.org/record/3758449}.

\bibitem[Etsebeth et~al.(2024)Etsebeth, Lochner, Walmsley, and
  Grespan]{etsebeth_2024_astronomaly_at_scale}
V.~Etsebeth, M.~Lochner, M.~Walmsley, and M.~Grespan.
\newblock \emph{Monthly Notices of the Royal Astronomical Society},
  529\penalty0 (1):\penalty0 732--747, 2024.
\newblock \doi{10.1093/mnras/stae496}.
\newblock URL \url{http://dx.doi.org/10.1093/mnras/stae496}.

\bibitem[Fu et~al.(2025)Fu, Xue, Zhu, Bhat, Wu, Zhang, Meyers, Tan, Yue, Niu,
  Meng, Wu, Fang, Zhou, and Jin]{fuImprovingPulsarSearch2025}
Q.~Fu et al.
\newblock \emph{Mon Not R Astron Soc}, page staf1952, Nov. 2025.
\newblock ISSN 0035-8711.
\newblock \doi{10.1093/mnras/staf1952}.
\newblock URL \url{https://doi.org/10.1093/mnras/staf1952}.

\bibitem[Gal and Ghahramani(2016)]{gal2016dropout}
Y.~Gal and Z.~Ghahramani.
\newblock In M.~F. Balcan and K.~Q. Weinberger, editors, \emph{Proceedings of
  The 33rd International Conference on Machine Learning}, volume~48 of
  \emph{Proceedings of Machine Learning Research}, pages 1050--1059, New York,
  New York, USA, 20--22 Jun 2016. PMLR.
\newblock URL \url{https://proceedings.mlr.press/v48/gal16.html}.

\bibitem[Gao et~al.(2025)Gao, Koopmans, Mertens, Munshi, Li, Brackenhoff,
  Ceccotti, Chege, Acharya, Ghara, Giri, Iliev, Mellema, and
  Zhang]{gaoExtractingEpochReionization2025a}
L.-Y. Gao et al.
\newblock \emph{ApJ}, 988\penalty0 (1):\penalty0 84, July 2025.
\newblock ISSN 0004-637X.
\newblock \doi{10.3847/1538-4357/ade2dc}.
\newblock URL \url{https://doi.org/10.3847/1538-4357/ade2dc}.

\bibitem[Goodfellow et~al.(2014)Goodfellow, Pouget-Abadie, Mirza, Xu,
  Warde-Farley, Ozair, Courville, and Bengio]{goodfellow2014generative}
I.~Goodfellow et al.
\newblock \emph{Advances in neural information processing systems}, 27, 2014.
\newblock \doi{10.48550/arXiv.1406.2661}.
\newblock URL \url{https://doi.org/10.48550/arXiv.1406.2661}.

\bibitem[Goodfellow et~al.(2016)Goodfellow, Bengio, and
  Courville]{Goodfellow-et-al-2016}
I.~Goodfellow, Y.~Bengio, and A.~Courville.
\newblock \emph{Deep Learning}.
\newblock MIT Press, 2016.
\newblock \url{http://www.deeplearningbook.org}.

\bibitem[Goodfellow et~al.(2020)Goodfellow, Pouget-Abadie, Mirza, Xu,
  Warde-Farley, Ozair, Courville, and Bengio]{goodfellow2020generative}
I.~Goodfellow et al.
\newblock \emph{Communications of the ACM}, 63\penalty0 (11):\penalty0
  139--144, 2020.
\newblock ISSN 1557-7317.
\newblock \doi{10.1145/3422622}.
\newblock URL \url{https://doi.org/10.1145/3422622}.

\bibitem[Gui et~al.(2021)Gui, Sun, Wen, Tao, and Ye]{gui2021review}
J.~Gui et al.
\newblock \emph{IEEE transactions on knowledge and data engineering},
  35\penalty0 (4):\penalty0 3313--3332, 2021.
\newblock ISSN 2326-3865.
\newblock \doi{10.1109/tkde.2021.3130191}.
\newblock URL \url{https://doi.org/10.1109/tkde.2021.3130191}.

\bibitem[Guidotti et~al.(2018)Guidotti, Monreale, Ruggieri, Turini, Giannotti,
  and Pedreschi]{guidotti2018survey}
R.~Guidotti et al.
\newblock \emph{ACM computing surveys (CSUR)}, 51\penalty0 (5):\penalty0 1--42,
  Aug. 2018.
\newblock ISSN 1557-7341.
\newblock \doi{10.1145/3236009}.
\newblock URL \url{https://doi.org/10.1145/3236009}.

\bibitem[Gustafsson et~al.(2024)Gustafsson, Br{\"u}ggen, and
  En{\ss}lin]{gustafssonSemiSupervisedRotationMeasure2024}
V.~Gustafsson, M.~Br{\"u}ggen, and T.~En{\ss}lin.
\newblock \emph{A\&A}, 692:\penalty0 A248, Dec. 2024.
\newblock ISSN 0004-6361, 1432-0746.
\newblock \doi{10.1051/0004-6361/202451265}.
\newblock URL \url{https://doi.org/10.1051/0004-6361/202451265}.

\bibitem[Hale and Tabatabaei(2026)]{Hale01.2026.SKA}
C.~Hale and F.~Tabatabaei.
\newblock In \emph{Advancing Astrophysics with the SKA -- II (AASKAII)}. 2026.
\newblock arXiv search: Report number AASKAII/Hale01.

\bibitem[Hancock et~al.(2012)Hancock, Murphy, Gaensler, Hopkins, and
  Curran]{hancock_2012_aegean_1}
P.~J. Hancock et al.
\newblock \emph{Monthly Notices of the Royal Astronomical Society},
  422\penalty0 (2):\penalty0 1812--1824, 05 2012.
\newblock ISSN 0035-8711.
\newblock \doi{10.1111/j.1365-2966.2012.20768.x}.
\newblock URL \url{https://doi.org/10.1111/j.1365-2966.2012.20768.x}.

\bibitem[Hancock et~al.(2018)Hancock, Trott, and
  Hurley-Walker]{hancock_2018_aegean_2}
P.~J. Hancock, C.~M. Trott, and N.~Hurley-Walker.
\newblock \emph{Publications of the Astronomical Society of Australia},
  35:\penalty0 e011, 2018.
\newblock \doi{10.1017/pasa.2018.3}.
\newblock URL \url{https://doi.org/10.1017/pasa.2018.3}.

\bibitem[Hardcastle et~al.(2026)Hardcastle, author2, author3, author4, and
  author5]{Hardcastle01.2026.SKA}
M.~J. Hardcastle et al.
\newblock In \emph{Advancing Astrophysics with the SKA -- II (AASKAII)}. 2026.
\newblock arXiv search: Report number AASKAII/Hardcastle01.

\bibitem[He et~al.(2016{\natexlab{a}})He, Zhang, Ren, and Sun]{resnet}
K.~He, X.~Zhang, S.~Ren, and J.~Sun.
\newblock In \emph{2016 IEEE Conference on Computer Vision and Pattern
  Recognition (CVPR)}, pages 770--778. IEEE, 2016{\natexlab{a}}.
\newblock \doi{10.1109/CVPR.2016.90}.
\newblock URL \url{https://doi.org/10.1109/CVPR.2016.90}.

\bibitem[He et~al.(2016{\natexlab{b}})He, Zhang, Ren, and Sun]{resnet2}
K.~He, X.~Zhang, S.~Ren, and J.~Sun.
\newblock \emph{Identity Mappings in Deep Residual Networks}, pages 630--645.
\newblock Springer International Publishing, Cham, 2016{\natexlab{b}}.
\newblock ISBN 9783319464930.
\newblock \doi{10.1007/978-3-319-46493-0_38}.
\newblock URL \url{https://doi.org/10.1007/978-3-319-46493-0_38}.

\bibitem[He et~al.(2018)He, Gkioxari, Dollár, and Girshick]{he2018maskrcnn}
K.~He, G.~Gkioxari, P.~Dollár, and R.~Girshick.
\newblock Mask r-cnn, Feb. 2018.
\newblock ISSN 1939-3539.
\newblock URL \url{https://doi.org/10.1109/tpami.2018.2844175}.

\bibitem[Heald et~al.(2020)Heald, Mao, Vacca, Akahori, Damas-Segovia, Gaensler,
  Hoeft, Agudo, Basu, Beck, Birkinshaw, Bonafede, Bourke, Bracco, Carretti,
  Feretti, Girart, Govoni, Green, Han, Haverkorn, Horellou, Johnston-Hollitt,
  Kothes, Landecker, Nikiel-Wroczyński, O'Sullivan, Padovani, Poidevin,
  Pratley, Regis, Riseley, Robishaw, Rudnick, Sobey, Stil, Sun, Sur, Taylor,
  Thomson, Eck, Vazza, West, and the SKA Magnetism Science
  Working~Group]{heald_2020_magnetism_science}
G.~Heald et al.
\newblock \emph{Galaxies}, 8\penalty0 (3):\penalty0 53, 2020.
\newblock \doi{10.3390/galaxies8030053}.
\newblock URL \url{http://dx.doi.org/10.3390/galaxies8030053}.

\bibitem[Hezaveh et~al.(2017)Hezaveh, Levasseur, and
  Marshall]{hezaveh_2017_fast_automated}
Y.~D. Hezaveh, L.~P. Levasseur, and P.~J. Marshall.
\newblock \emph{Nature}, 548\penalty0 (7669):\penalty0 555--557, 2017.
\newblock \doi{10.1038/nature23463}.
\newblock URL \url{http://dx.doi.org/10.1038/nature23463}.

\bibitem[Ho et~al.(2020)Ho, Jain, and Abbeel]{ho2020denoising}
J.~Ho, A.~Jain, and P.~Abbeel.
\newblock \emph{Advances in neural information processing systems},
  33:\penalty0 6840--6851, 2020.
\newblock \doi{10.48550/arXiv.2006.11239}.
\newblock URL \url{https://doi.org/10.48550/arXiv.2006.11239}.

\bibitem[Holderrieth and
  Erives(2025)]{holderrieth2025introductionflowmatchingdiffusion}
P.~Holderrieth and E.~Erives.
\newblock An introduction to flow matching and diffusion models, 2025.
\newblock URL \url{https://arxiv.org/abs/2506.02070}.

\bibitem[Hota et~al.(2026)Hota, author2, author3, author4, and
  author5]{Hota01.2026.SKA}
A.~Hota et al.
\newblock In \emph{Advancing Astrophysics with the SKA -- II (AASKAII)}. 2026.
\newblock arXiv search: Report number AASKAII/Hota01.

\bibitem[Hu(2025)]{hu_2025_characterizing_magnetic}
Y.~Hu.
\newblock \emph{The Astrophysical Journal}, 990\penalty0 (1):\penalty0 76,
  2025.
\newblock \doi{10.3847/1538-4357/adf32e}.
\newblock URL \url{http://dx.doi.org/10.3847/1538-4357/adf32e}.

\bibitem[Hu and Lazarian(2024)]{hu_2024_probing_three}
Y.~Hu and A.~Lazarian.
\newblock \emph{The Astrophysical Journal}, 975\penalty0 (1):\penalty0 66,
  2024.
\newblock \doi{10.3847/1538-4357/ad7950}.
\newblock URL \url{http://dx.doi.org/10.3847/1538-4357/ad7950}.

\bibitem[Hu and Lazarian(2025)]{hu_2025_probing_three}
Y.~Hu and A.~Lazarian.
\newblock \emph{The Astrophysical Journal}, 981\penalty0 (1):\penalty0 58,
  2025.
\newblock \doi{10.3847/1538-4357/adaf97}.
\newblock URL \url{http://dx.doi.org/10.3847/1538-4357/adaf97}.

\bibitem[Huyen(2022)]{huyen2022designing}
C.~Huyen.
\newblock \emph{Designing machine learning systems}.
\newblock " O'Reilly Media, Inc.", [S.l.], 2022.
\newblock ISBN 1098107969.

\bibitem[Isola et~al.(2017)Isola, Zhu, Zhou, and Efros]{isola2017image}
P.~Isola, J.-Y. Zhu, T.~Zhou, and A.~A. Efros.
\newblock In \emph{Proceedings of the IEEE conference on computer vision and
  pattern recognition}, pages 1125--1134. IEEE, 2017.
\newblock \doi{10.1109/cvpr.2017.632}.
\newblock URL \url{https://doi.org/10.1109/cvpr.2017.632}.

\bibitem[Kairouz et~al.(2021)Kairouz, McMahan, Avent, Bellet, Bennis,
  Nitin~Bhagoji, Bonawitz, Charles, Cormode, Cummings, D’Oliveira, Eichner,
  El~Rouayheb, Evans, Gardner, Garrett, Gascón, Ghazi, Gibbons, Gruteser,
  Harchaoui, He, He, Huo, Hutchinson, Hsu, Jaggi, Javidi, Joshi, Khodak,
  Konecný, Korolova, Koushanfar, Koyejo, Lepoint, Liu, Mittal, Mohri, Nock,
  Özgür, Pagh, Qi, Ramage, Raskar, Raykova, Song, Song, Stich, Sun, Suresh,
  Tramèr, Vepakomma, Wang, Xiong, Xu, Yang, Yu, Yu, and
  Zhao]{kairouz2021advancesopenproblemsfederated}
P.~Kairouz et al.
\newblock Advances and open problems in federated learning, 2021.
\newblock URL \url{https://doi.org/10.1561/9781680837896}.

\bibitem[Karniadakis et~al.(2021)Karniadakis, Kevrekidis, Lu, Perdikaris, Wang,
  and Yang]{karniadakis2021physics}
G.~E. Karniadakis et al.
\newblock \emph{Nature Reviews Physics}, 3\penalty0 (6):\penalty0 422--440,
  2021.
\newblock \doi{10.1038/s42254-021-00314-5}.
\newblock URL \url{https://doi.org/10.1038/s42254-021-00314-5}.

\bibitem[Karras et~al.(2022)Karras, Aittala, Aila, and
  Laine]{karras2022elucidating}
T.~Karras, M.~Aittala, T.~Aila, and S.~Laine.
\newblock \emph{Advances in Neural Information Processing Systems},
  35:\penalty0 26565--26577, 2022.
\newblock \doi{10.52202/068431-1926}.
\newblock URL \url{https://doi.org/10.52202/068431-1926}.

\bibitem[Keane et~al.(2026)Keane, author2, author3, author4, and
  author5]{Keane01.2026.SKA}
E.~F. Keane et al.
\newblock In \emph{Advancing Astrophysics with the SKA -- II (AASKAII)}. 2026.
\newblock arXiv search: Report number AASKAII/Keane01.

\bibitem[Kingma and Welling(2013)]{kingma2013auto}
D.~P. Kingma and M.~Welling.
\newblock \emph{arXiv preprint arXiv:1312.6114}, 2013.
\newblock \doi{10.48550/arXiv.1312.6114}.
\newblock URL \url{https://doi.org/10.48550/arXiv.1312.6114}.

\bibitem[Kirillov et~al.(2023)Kirillov, Mintun, Ravi, Mao, Rolland, Gustafson,
  Xiao, Whitehead, Berg, Lo, et~al.]{kirillov2023segment}
A.~Kirillov et al.
\newblock In \emph{Proceedings of the IEEE/CVF international conference on
  computer vision}, pages 4015--4026. IEEE, Oct. 2023.
\newblock \doi{10.1109/iccv51070.2023.00371}.
\newblock URL \url{https://doi.org/10.1109/iccv51070.2023.00371}.

\bibitem[Kirk et~al.(2024)Kirk, Rau, and
  Ramyaa]{kirk_2024_reinforcement_learning}
B.~M. Kirk, U.~Rau, and R.~Ramyaa.
\newblock \emph{The Astronomical Journal}, 169\penalty0 (1):\penalty0 43, 2024.
\newblock \doi{10.3847/1538-3881/ad88f6}.
\newblock URL \url{http://dx.doi.org/10.3847/1538-3881/ad88f6}.

\bibitem[Kobyzev et~al.(2020)Kobyzev, Prince, and
  Brubaker]{kobyzev2020normalizing}
I.~Kobyzev, S.~J. Prince, and M.~A. Brubaker.
\newblock \emph{IEEE transactions on pattern analysis and machine
  intelligence}, 43\penalty0 (11):\penalty0 3964--3979, Nov. 2020.
\newblock ISSN 1939-3539.
\newblock \doi{10.1109/tpami.2020.2992934}.
\newblock URL \url{https://doi.org/10.1109/tpami.2020.2992934}.

\bibitem[Korber et~al.(2023)Korber, Bianco, Tolley, and
  Kneib]{korber_2023_pinion}
D.~Korber, M.~Bianco, E.~Tolley, and J.-P. Kneib.
\newblock \emph{Monthly Notices of the Royal Astronomical Society},
  521\penalty0 (1):\penalty0 902--915, 2023.
\newblock \doi{10.1093/mnras/stad615}.
\newblock URL \url{http://dx.doi.org/10.1093/mnras/stad615}.

\bibitem[Krizhevsky et~al.(2017)Krizhevsky, Sutskever, and Hinton]{alexnet}
A.~Krizhevsky, I.~Sutskever, and G.~E. Hinton.
\newblock \emph{Commun. ACM}, 60\penalty0 (6):\penalty0 84–90, May 2017.
\newblock ISSN 0001-0782.
\newblock \doi{10.1145/3065386}.
\newblock URL \url{https://doi.org/10.1145/3065386}.

\bibitem[Kuiper et~al.(2025)Kuiper, Contardo, Huppenkothen, and
  Hessels]{kuiper_2025_representation_learning}
D.~Kuiper, G.~Contardo, D.~Huppenkothen, and J.~W.~T. Hessels.
\newblock \emph{Monthly Notices of the Royal Astronomical Society},
  538\penalty0 (1):\penalty0 408--425, 03 2025.
\newblock ISSN 0035-8711.
\newblock \doi{10.1093/mnras/staf306}.
\newblock URL \url{https://doi.org/10.1093/mnras/staf306}.

\bibitem[La~Plante and Ntampaka(2019)]{laplanteMachineLearningApplied2019}
P.~La~Plante and M.~Ntampaka.
\newblock \emph{ApJ}, 880\penalty0 (2):\penalty0 110, July 2019.
\newblock ISSN 0004-637X.
\newblock \doi{10.3847/1538-4357/ab2983}.
\newblock URL \url{https://doi.org/10.3847/1538-4357/ab2983}.

\bibitem[Lagos et~al.(2026)Lagos, author2, author3, author4, and
  author5]{Lagos01.2026.SKA}
C.~D. Lagos et al.
\newblock In \emph{Advancing Astrophysics with the SKA -- II (AASKAII)}. 2026.
\newblock arXiv search: Report number AASKAII/Lagos01.

\bibitem[Lastufka et~al.(2024)Lastufka, Bait, Taran, Drozdova, Kinakh, Piras,
  Audard, {Dessauges-Zavadsky}, Holotyak, Schaerer, and
  Voloshynovskiy]{lastufkaSelfSupervisedLearningMeerKAT2024}
E.~Lastufka et al.
\newblock \emph{A\&A}, 690:\penalty0 A310, Oct. 2024.
\newblock ISSN 0004-6361, 1432-0746.
\newblock \doi{10.1051/0004-6361/202449964}.
\newblock URL \url{https://doi.org/10.1051/0004-6361/202449964}.

\bibitem[Lastufka et~al.(2025)Lastufka, Bait, Drozdova, Kinakh, Piras, Audard,
  Dessauges-Zavadsky, Holotyak, Schaerer, and
  Voloshynovskiy]{lastufka_2025_examining_vision}
E.~Lastufka et al.
\newblock \emph{Astronomy \& Astrophysics}, 703:\penalty0 A217, 2025.
\newblock \doi{10.1051/0004-6361/202553691}.
\newblock URL \url{http://dx.doi.org/10.1051/0004-6361/202553691}.

\bibitem[Lecun et~al.(1998)Lecun, Bottou, Bengio, and Haffner]{lecun1998}
Y.~Lecun, L.~Bottou, Y.~Bengio, and P.~Haffner.
\newblock \emph{Proceedings of the IEEE}, 86\penalty0 (11):\penalty0
  2278--2324, 1998.
\newblock ISSN 0018-9219.
\newblock \doi{10.1109/5.726791}.
\newblock URL \url{https://doi.org/10.1109/5.726791}.

\bibitem[LeCun et~al.(2006)LeCun, Chopra, Hadsell, Ranzato, and
  Huang]{lecun2006tutorial}
Y.~LeCun et al.
\newblock \emph{{A tutorial on energy-based learning}}.
\newblock MIT Press, 2006.
\newblock URL
  \url{https://nyuscholars.nyu.edu/en/publications/a-tutorial-on-energy-based-learning}.

\bibitem[LeCun et~al.(2015)LeCun, Bengio, and Hinton]{lecun2015deep}
Y.~LeCun, Y.~Bengio, and G.~Hinton.
\newblock \emph{nature}, 521\penalty0 (7553):\penalty0 436--444, May 2015.
\newblock ISSN 1476-4687.
\newblock \doi{10.1038/nature14539}.
\newblock URL \url{https://doi.org/10.1038/nature14539}.

\bibitem[Legin et~al.(2021)Legin, Hezaveh, Levasseur, and
  Wandelt]{legin_2021_simulation_based}
R.~Legin, Y.~Hezaveh, L.~P. Levasseur, and B.~Wandelt.
\newblock \emph{CoRR}, 2021.
\newblock URL \url{http://arxiv.org/abs/2112.05278v2}.

\bibitem[Li et~al.(2019)Li, Xu, Ma, Zhu, Hu, Zhu, Gu, Shan, Zhu, and
  Wu]{liSeparatingEoRSignal2019}
W.~Li et al.
\newblock \emph{Mon Not R Astron Soc}, 485\penalty0 (2):\penalty0 2628--2637,
  May 2019.
\newblock ISSN 0035-8711.
\newblock \doi{10.1093/mnras/stz582}.
\newblock URL \url{https://doi.org/10.1093/mnras/stz582}.

\bibitem[Li et~al.(2021)Li, Yu, Xiao, Long, and Cui]{li_2021_detection_radio}
Z.~Li et al.
\newblock \emph{Astronomy and Computing}, 36:\penalty0 100482, 2021.
\newblock ISSN 2213-1337.
\newblock \doi{https://doi.org/10.1016/j.ascom.2021.100482}.
\newblock URL
  \url{https://www.sciencedirect.com/science/article/pii/S2213133721000366}.

\bibitem[Liaudat et~al.(2023)Liaudat, Mars, Price, Pereyra, Betcke, and
  McEwen]{liaudat_2023_scalable_bayesian}
T.~I. Liaudat et al.
\newblock \emph{CoRR}, 2023.
\newblock URL \url{http://arxiv.org/abs/2312.00125v3}.

\bibitem[Lipman et~al.(2023)Lipman, Chen, Ben-Hamu, Nickel, and
  Le]{lipman2023flowmatchinggenerativemodeling}
Y.~Lipman et al.
\newblock Flow matching for generative modeling, 2023.
\newblock URL \url{https://doi.org/10.48550/arXiv.2210.02747}.

\bibitem[Lipman et~al.(2024)Lipman, Havasi, Holderrieth, Shaul, Le, Karrer,
  Chen, Lopez-Paz, Ben-Hamu, and Gat]{lipman2024flowmatchingguidecode}
Y.~Lipman et al.
\newblock Flow matching guide and code, 2024.
\newblock URL \url{https://doi.org/10.48550/arXiv.2412.06264}.

\bibitem[Liu et~al.(2022)Liu, Gong, and Liu]{liu2022flowstraightfastlearning}
X.~Liu, C.~Gong, and Q.~Liu.
\newblock Flow straight and fast: Learning to generate and transfer data with
  rectified flow, 2022.
\newblock URL \url{https://doi.org/10.48550/arXiv.2209.03003}.

\bibitem[Liu et~al.(2025{\natexlab{a}})Liu, Acedo, and
  Sims]{liuBayesianModelComparison2025}
Y.~Liu, E.~d.~L. Acedo, and P.~Sims.
\newblock Bayesian model comparison and validation with {{Gaussian Process
  Regression}} for interferometric 21-cm signal recovery, Nov.
  2025{\natexlab{a}}.
\newblock URL \url{https://doi.org/10.48550/arXiv.2511.10499}.

\bibitem[Liu et~al.(2025{\natexlab{b}})Liu, Tu, Li, Miao, Lin, and
  Zhao]{liu_2025_yolox_ls}
Y.~Liu et al.
\newblock \emph{The Astrophysical Journal}, 978\penalty0 (2):\penalty0 151,
  2025{\natexlab{b}}.
\newblock \doi{10.3847/1538-4357/ad97ba}.
\newblock URL \url{http://dx.doi.org/10.3847/1538-4357/ad97ba}.

\bibitem[Loi et~al.(2026)Loi, author2, author3, author4, and
  author5]{Loi01.2026.SKA}
F.~Loi et al.
\newblock In \emph{Advancing Astrophysics with the SKA -- II (AASKAII)}. 2026.
\newblock arXiv search: Report number AASKAII/Loi01.

\bibitem[Ma et~al.(2026)Ma, author2, author3, author4, and
  author5]{Ma01.2026.SKA}
Y.~K. Ma et al.
\newblock In \emph{Advancing Astrophysics with the SKA -- II (AASKAII)}. 2026.
\newblock arXiv search: Report number AASKAII/Ma01.

\bibitem[MacKay(1992)]{mackay1992}
D.~J.~C. MacKay.
\newblock \emph{Neural Computation}, 4\penalty0 (3):\penalty0 448--472, May
  1992.
\newblock ISSN 1530-888X.
\newblock \doi{10.1162/neco.1992.4.3.448}.
\newblock URL \url{https://doi.org/10.1162/neco.1992.4.3.448}.

\bibitem[Mahida et~al.(2025)Mahida, Yadav, Majumdar, Noble, Murmu, Dasgupta,
  Dutta, Tiwari, and Shaw]{mahidaANNBNNInferring2025}
Y.~Mahida et al.
\newblock \emph{Journal of Cosmology and Astroparticle Physics}, 2025\penalty0
  (12):\penalty0 055, Dec. 2025.
\newblock ISSN 1475-7516.
\newblock \doi{10.1088/1475-7516/2025/12/055}.
\newblock URL \url{https://doi.org/10.1088/1475-7516/2025/12/055}.

\bibitem[Mangena et~al.(2020)Mangena, Hassan, and
  Santos]{mangenaConstrainingReionizationHistory2020}
T.~Mangena, S.~Hassan, and M.~G. Santos.
\newblock \emph{Mon Not R Astron Soc}, 494\penalty0 (1):\penalty0 600--606, May
  2020.
\newblock ISSN 0035-8711.
\newblock \doi{10.1093/mnras/staa750}.
\newblock URL \url{https://doi.org/10.1093/mnras/staa750}.

\bibitem[Margalef-Bentabol et~al.(2020)Margalef-Bentabol, Huertas-Company,
  Charnock, Margalef-Bentabol, Bernardi, Dubois, Storey-Fisher, and
  Zanisi]{margalef-bentabol_2020_detecting_outliers}
B.~Margalef-Bentabol et al.
\newblock \emph{Monthly Notices of the Royal Astronomical Society},
  496\penalty0 (2):\penalty0 2346--2361, 08 2020.
\newblock ISSN 0035-8711.
\newblock \doi{10.1093/mnras/staa1647}.
\newblock URL \url{https://doi.org/10.1093/mnras/staa1647}.

\bibitem[Mart{\'i}nez et~al.(2024)Mart{\'i}nez, P{\'e}rez, and
  Br{\"u}ggen]{martinezSimulatingImagesRadio2024}
T.~V. Mart{\'i}nez, N.~B. P{\'e}rez, and M.~Br{\"u}ggen.
\newblock \emph{A\&A}, 691:\penalty0 A360, Nov. 2024.
\newblock ISSN 0004-6361, 1432-0746.
\newblock \doi{10.1051/0004-6361/202451429}.
\newblock URL \url{https://doi.org/10.1051/0004-6361/202451429}.

\bibitem[{McKean} et~al.(2015){McKean}, {Jackson}, {Vegetti}, {Rybak},
  {Serjeant}, {Koopmans}, {Metcalf}, {Fassnacht}, {Marshall}, and
  {Pandey-Pommier}]{mckean_2015_strong_gravitational}
J.~{McKean} et al.
\newblock In \emph{Advancing Astrophysics with the Square Kilometre Array
  (AASKA14)}, page~84, Apr. 2015.
\newblock \doi{10.22323/1.215.0084}.
\newblock URL \url{https://doi.org/10.22323/1.215.0084}.

\bibitem[McMahan et~al.(2023)McMahan, Moore, Ramage, Hampson, and
  y~Arcas]{mcmahan2016communicationefficientlearningdeepnetworks}
H.~B. McMahan et al.
\newblock Communication-efficient learning of deep networks from decentralized
  data, 2023.
\newblock URL \url{https://doi.org/10.48550/arXiv.1602.05629}.

\bibitem[Mertens et~al.(2025)Mertens, Mevius, Koopmans, Offringa, Zaroubi,
  Acharya, Brackenhoff, Ceccotti, Chapman, Chege, Ciardi, Ghara, Ghosh, Giri,
  Hothi, H{\"o}fer, Iliev, Jeli{\'c}, Ma, Mellema, Munshi, Pandey, and
  Yatawatta]{mertensDeeperMultiredshiftUpper2025a}
F.~G. Mertens et al.
\newblock \emph{A\&A}, 698:\penalty0 A186, June 2025.
\newblock ISSN 0004-6361, 1432-0746.
\newblock \doi{10.1051/0004-6361/202554158}.
\newblock URL \url{https://doi.org/10.1051/0004-6361/202554158}.

\bibitem[Mesinger et~al.(2010)Mesinger, Furlanetto, and
  Cen]{mesinger_2010_21cmfast}
A.~Mesinger, S.~Furlanetto, and R.~Cen.
\newblock \emph{Monthly Notices of the Royal Astronomical Society},
  411\penalty0 (2):\penalty0 955--972, 2010.
\newblock \doi{10.1111/j.1365-2966.2010.17731.x}.
\newblock URL \url{http://dx.doi.org/10.1111/j.1365-2966.2010.17731.x}.

\bibitem[Miller-Jones et~al.(2026)Miller-Jones, Rajwade, Woudt, and
  Hessels]{Miller-Jones01.2026.SKA}
J.~Miller-Jones, K.~Rajwade, P.~A. Woudt, and J.~W. Hessels.
\newblock In \emph{Advancing Astrophysics with the SKA -- II (AASKAII)}. 2026.
\newblock arXiv search: Report number AASKAII/Miller-Jones01.

\bibitem[Mitchell(1997)]{mitchell1997machine}
T.~M. Mitchell.
\newblock \emph{Machine learning}.
\newblock McGraw-Hill, New York, 1997.
\newblock URL \url{https://www.cs.cmu.edu/~tom/mlbook.html}.

\bibitem[Mnih et~al.(2015)Mnih, Kavukcuoglu, Silver, Rusu, Veness, Bellemare,
  Graves, Riedmiller, Fidjeland, Ostrovski, Petersen, Beattie, Sadik,
  Antonoglou, King, Kumaran, Wierstra, Legg, and Hassabis]{mnih2015human}
V.~Mnih et al.
\newblock \emph{Nature}, 518\penalty0 (7540):\penalty0 529--533, Feb. 2015.
\newblock ISSN 1476-4687.
\newblock \doi{10.1038/nature14236}.
\newblock URL \url{https://doi.org/10.1038/nature14236}.

\bibitem[{Mohan} and {Rafferty}(2015)]{mohan_2015_pybdsf}
N.~{Mohan} and D.~{Rafferty}.
\newblock {PyBDSF: Python Blob Detection and Source Finder}.
\newblock Astrophysics Source Code Library, record ascl:1502.007, Feb. 2015.

\bibitem[Molnar(2025)]{molnar2020interpretable}
C.~Molnar.
\newblock \emph{Interpretable Machine Learning}.
\newblock 3 edition, 2025.
\newblock ISBN 978-3-911578-03-5.
\newblock URL \url{https://christophm.github.io/interpretable-ml-book}.

\bibitem[Morningstar et~al.(2018)Morningstar, Hezaveh, Levasseur, Blandford,
  Marshall, Putzky, and Wechsler]{morningstar_2018_analyzing_interferometric}
W.~R. Morningstar et al.
\newblock \emph{CoRR}, 2018.
\newblock URL \url{http://arxiv.org/abs/1808.00011v1}.

\bibitem[Morningstar et~al.(2019)Morningstar, Levasseur, Hezaveh, Blandford,
  Marshall, Putzky, Rueter, Wechsler, and
  Welling]{morningstar_2019_data_driven}
W.~R. Morningstar et al.
\newblock \emph{The Astrophysical Journal}, 883\penalty0 (1):\penalty0 14,
  2019.
\newblock \doi{10.3847/1538-4357/ab35d7}.
\newblock URL \url{http://dx.doi.org/10.3847/1538-4357/ab35d7}.

\bibitem[Mostert et~al.(2022)Mostert, Duncan, Alegre, R{\"o}ttgering, Williams,
  Best, Hardcastle, and Morganti]{mostert_2022_radio_source_component}
R.~I.~J. Mostert et al.
\newblock \emph{Astronomy \& Astrophysics}, 668:\penalty0 A28, 2022.
\newblock \doi{10.1051/0004-6361/202243478}.
\newblock URL \url{http://dx.doi.org/10.1051/0004-6361/202243478}.

\bibitem[Murphy(2025)]{murphy2024reinforcementlearningoverview}
K.~Murphy.
\newblock Reinforcement learning: An overview, 2025.
\newblock URL \url{https://arxiv.org/abs/2412.05265}.

\bibitem[Ndiritu et~al.(2021)Ndiritu, Scaife, Tabb, C{\'a}rcamo, and
  Hanson]{ndiritu_2021_gaussian_process}
S.~W. Ndiritu et al.
\newblock \emph{Monthly Notices of the Royal Astronomical Society},
  502\penalty0 (4):\penalty0 5839--5853, 2021.
\newblock \doi{10.1093/mnras/stab379}.
\newblock URL \url{http://dx.doi.org/10.1093/mnras/stab379}.

\bibitem[Neal(2012)]{neal2012bayesian}
R.~M. Neal.
\newblock \emph{Bayesian learning for neural networks}, volume 118.
\newblock Springer Science \& Business Media, 2012.

\bibitem[Neutsch et~al.(2022)Neutsch, Heneka, and
  Br{\"u}ggen]{neutschInferringAstrophysicsDark2022}
S.~Neutsch, C.~Heneka, and M.~Br{\"u}ggen.
\newblock \emph{Mon Not R Astron Soc}, 511\penalty0 (3):\penalty0 3446--3462,
  Apr. 2022.
\newblock ISSN 0035-8711.
\newblock \doi{10.1093/mnras/stac218}.
\newblock URL \url{https://doi.org/10.1093/mnras/stac218}.

\bibitem[Nguyen et~al.(2025)Nguyen, Tang, Alger, Marchal, Muller, Ong, and
  {McClure-Griffiths}]{nguyenTPCNetRepresentationLearning2025}
H.~Nguyen et al.
\newblock \emph{Mon Not R Astron Soc}, 536\penalty0 (1):\penalty0 962--987,
  Jan. 2025.
\newblock ISSN 0035-8711.
\newblock \doi{10.1093/mnras/stae2631}.
\newblock URL \url{https:/doi.org/10.1093/mnras/stae2631}.

\bibitem[Nightingale et~al.(2018)Nightingale, Dye, and
  Massey]{nightingale_2018_autolens}
J.~W. Nightingale, S.~Dye, and R.~J. Massey.
\newblock \emph{Monthly Notices of the Royal Astronomical Society},
  478\penalty0 (4):\penalty0 4738--4784, 2018.
\newblock \doi{10.1093/mnras/sty1264}.
\newblock URL \url{http://dx.doi.org/10.1093/mnras/sty1264}.

\bibitem[Oppliger et~al.(2024)Oppliger, Denner, K{\"u}spert, Frison, Wang,
  Morawietz, Ivashko, Dippel, von Zimmermann, Biało, Martinelli, Fauqu{\'e},
  Choi, Garcia-Fernandez, Zhou, Christensen, Kurosawa, Momono, Oda, Natterer,
  Fischer, Neupert, and Chang]{oppliger_2024_weak_signal}
J.~Oppliger et al.
\newblock \emph{Nature Machine Intelligence}, 6\penalty0 (2):\penalty0
  180--186, 2024.
\newblock \doi{10.1038/s42256-024-00790-1}.
\newblock URL \url{http://dx.doi.org/10.1038/s42256-024-00790-1}.

\bibitem[O'Sullivan et~al.(2026)O'Sullivan, author2, author3, author4, and
  author5]{OSullivan01.2026.SKA}
S.~P. O'Sullivan et al.
\newblock In \emph{Advancing Astrophysics with the SKA -- II (AASKAII)}. 2026.
\newblock arXiv search: Report number AASKAII/OSullivan01.

\bibitem[Pal et~al.(2026)Pal, author2, author3, author4, and
  author5]{SabyasachiPal01.2026.SKA}
S.~Pal et al.
\newblock In \emph{Advancing Astrophysics with the SKA -- II (AASKAII)}. 2026.
\newblock arXiv search: Report number AASKAII/SabyasachiPal01.

\bibitem[Pang et~al.(2021)Pang, Shen, Cao, and Hengel]{pang2021ano}
G.~Pang, C.~Shen, L.~Cao, and A.~V.~D. Hengel.
\newblock \emph{ACM Comput. Surv.}, 54\penalty0 (2), Mar. 2021.
\newblock ISSN 0360-0300.
\newblock \doi{10.1145/3439950}.
\newblock URL \url{https://doi.org/10.1145/3439950}.

\bibitem[Papamakarios et~al.(2021)Papamakarios, Nalisnick, Rezende, Mohamed,
  and Lakshminarayanan]{papamakarios2021normalizing}
G.~Papamakarios et al.
\newblock \emph{The Journal of Machine Learning Research}, 22\penalty0
  (1):\penalty0 2617--2680, 2021.
\newblock URL \url{http://jmlr.org/papers/v22/19-1028.html}.

\bibitem[Park et~al.(2021)Park, Ko, Huh, and Kim]{park2021review}
S.-W. Park, J.-S. Ko, J.-H. Huh, and J.-C. Kim.
\newblock \emph{Electronics}, 10\penalty0 (10):\penalty0 1216, 2021.
\newblock \doi{10.3390/electronics10101216}.
\newblock URL \url{https://doi.org/10.3390/electronics10101216}.

\bibitem[Parker et~al.(2024)Parker, Lanusse, Golkar, Sarra, Cranmer, Bietti,
  Eickenberg, Krawezik, McCabe, Morel, Ohana, Pettee, Régaldo-Saint Blancard,
  Cho, Ho, and Collaboration]{parker_2024_astroclip}
L.~Parker et al.
\newblock \emph{Monthly Notices of the Royal Astronomical Society},
  531\penalty0 (4):\penalty0 4990--5011, 07 2024.
\newblock ISSN 0035-8711.
\newblock \doi{10.1093/mnras/stae1450}.
\newblock URL \url{https://doi.org/10.1093/mnras/stae1450}.

\bibitem[Pattle et~al.(2022)Pattle, Fissel, Tahani, Liu, and
  Ntormousi]{pattle_2022_magnetic_fields}
K.~Pattle et al.
\newblock \emph{CoRR}, 2022.
\newblock URL \url{http://arxiv.org/abs/2203.11179v1}.

\bibitem[Peek and Burkhart(2019)]{peek_2019_do_androids}
J.~E.~G. Peek and B.~Burkhart.
\newblock \emph{The Astrophysical Journal Letters}, 882\penalty0 (1):\penalty0
  L12, 2019.
\newblock \doi{10.3847/2041-8213/ab3a9e}.
\newblock URL \url{http://dx.doi.org/10.3847/2041-8213/ab3a9e}.

\bibitem[Pietschke et~al.(2025)Pietschke, Heneka, Schlenker, Ore, and
  Schosser]{pietschkeDirectReconstructionReionization2025}
Y.~Pietschke et al.
\newblock \emph{J. Cosmol. Astropart. Phys.}, 2025\penalty0 (10):\penalty0 039,
  Oct. 2025.
\newblock ISSN 1475-7516.
\newblock \doi{10.1088/1475-7516/2025/10/039}.
\newblock URL \url{https://doi.org/10.1088/1475-7516/2025/10/039}.

\bibitem[Piras et~al.(2024)Piras, Polanska, Mancini, Price, and
  McEwen]{piras_2024_future_cosmological}
D.~Piras et al.
\newblock \emph{The Open Journal of Astrophysics}, 7, 2024.
\newblock \doi{10.33232/001c.123368}.
\newblock URL \url{http://dx.doi.org/10.33232/001c.123368}.

\bibitem[Prandoni and Seymour(2015)]{prandoni_2015_revealing_physics}
I.~Prandoni and N.~Seymour.
\newblock In \emph{Proceedings of Advancing Astrophysics with the Square
  Kilometre Array — PoS(AASKA14)}, AASKA14, page~67. Sissa Medialab, Apr.
  2015.
\newblock \doi{10.22323/1.215.0067}.
\newblock URL \url{https://doi.org/10.22323/1.215.0067}.

\bibitem[Prandoni et~al.(2026)Prandoni, author2, author3, author4, and
  author5]{Prandoni01.2026.SKA}
I.~Prandoni et al.
\newblock In \emph{Advancing Astrophysics with the SKA -- II (AASKAII)}. 2026.
\newblock arXiv search: Report number AASKAII/Prandoni01.

\bibitem[Prelogović and Mesinger(2023)]{prelogovic_2023_exploring_likelihood}
D.~Prelogović and A.~Mesinger.
\newblock \emph{Monthly Notices of the Royal Astronomical Society},
  524\penalty0 (3):\penalty0 4239--4255, 09 2023.
\newblock ISSN 0035-8711.
\newblock \doi{10.1093/mnras/stad2027}.
\newblock URL \url{https://doi.org/10.1093/mnras/stad2027}.

\bibitem[Pritchard et~al.(2025)Pritchard, Wicenec, Bennamoun, and
  Dodson]{pritchard_2025_advancing_rfi}
N.~J. Pritchard, A.~Wicenec, M.~Bennamoun, and R.~Dodson.
\newblock \emph{CoRR}, 2025.
\newblock URL \url{http://arxiv.org/abs/2504.09796v1}.

\bibitem[Qiu et~al.(2026)Qiu, author2, author3, author4, and
  author5]{Qiu01.2026.SKA}
H.~Qiu et al.
\newblock In \emph{Advancing Astrophysics with the SKA -- II (AASKAII)}. 2026.
\newblock arXiv search: Report number AASKAII/Qiu01.

\bibitem[Radford et~al.(2015)Radford, Metz, and
  Chintala]{radford2015unsupervised}
A.~Radford, L.~Metz, and S.~Chintala.
\newblock 2015.
\newblock \doi{10.48550/arXiv.1511.06434}.
\newblock URL \url{https://doi.org/10.48550/arXiv.1511.06434}.

\bibitem[Radford et~al.(2021)Radford, Kim, Hallacy, Ramesh, Goh, Agarwal,
  Sastry, Askell, Mishkin, Clark, Krueger, and
  Sutskever]{radford2021learningtransferablevisualmodels}
A.~Radford et al.
\newblock Learning transferable visual models from natural language
  supervision, 2021.
\newblock URL \url{https://arxiv.org/abs/2103.00020}.

\bibitem[Raissi et~al.(2017{\natexlab{a}})Raissi, Perdikaris, and
  Karniadakis]{raissi2017physicsinformeddeeplearning}
M.~Raissi, P.~Perdikaris, and G.~E. Karniadakis.
\newblock Physics informed deep learning (part i): Data-driven solutions of
  nonlinear partial differential equations, 2017{\natexlab{a}}.
\newblock URL \url{https://arxiv.org/abs/1711.10561}.

\bibitem[Raissi et~al.(2017{\natexlab{b}})Raissi, Perdikaris, and
  Karniadakis]{raissi2017physicsinformeddeeplearning2}
M.~Raissi, P.~Perdikaris, and G.~E. Karniadakis.
\newblock Physics informed deep learning (part ii): Data-driven discovery of
  nonlinear partial differential equations, 2017{\natexlab{b}}.
\newblock URL \url{https://arxiv.org/abs/1711.10566}.

\bibitem[Raissi et~al.(2019)Raissi, Perdikaris, and Karniadakis]{RAISSI2019686}
M.~Raissi, P.~Perdikaris, and G.~Karniadakis.
\newblock \emph{Journal of Computational Physics}, 378:\penalty0 686--707,
  2019.
\newblock ISSN 0021-9991.
\newblock \doi{https://doi.org/10.1016/j.jcp.2018.10.045}.
\newblock URL
  \url{https://www.sciencedirect.com/science/article/pii/S0021999118307125}.

\bibitem[Redmon et~al.(2016)Redmon, Divvala, Girshick, and
  Farhadi]{redmon2016you}
J.~Redmon, S.~Divvala, R.~Girshick, and A.~Farhadi.
\newblock In \emph{Proceedings of the IEEE conference on computer vision and
  pattern recognition}, pages 779--788, 2016.
\newblock \doi{10.1109/cvpr.2016.91}.
\newblock URL \url{https://doi.org/10.1109/cvpr.2016.91}.

\bibitem[Remy et~al.(2023)Remy, Lanusse, Jeffrey, Liu, Starck, Osato, and
  Schrabback]{remy_2023_probabilistic_mass}
B.~Remy et al.
\newblock \emph{Astronomy \& Astrophysics}, 672:\penalty0 A51, 2023.
\newblock \doi{10.1051/0004-6361/202243054}.
\newblock URL \url{http://dx.doi.org/10.1051/0004-6361/202243054}.

\bibitem[Ren et~al.(2016)Ren, He, Girshick, and
  Sun]{ren2016fasterrcnnrealtimeobject}
S.~Ren, K.~He, R.~Girshick, and J.~Sun.
\newblock Faster r-cnn: Towards real-time object detection with region proposal
  networks, 2016.
\newblock URL \url{https://arxiv.org/abs/1506.01497}.

\bibitem[Rezaei et~al.(2022)Rezaei, McKean, Biehl, de~Roo1, and
  Lafontaine]{rezaeiMachineLearningBased2022}
S.~Rezaei et al.
\newblock \emph{Monthly Notices of the Royal Astronomical Society},
  517\penalty0 (1):\penalty0 1156--1170, Oct. 2022.
\newblock ISSN 0035-8711, 1365-2966.
\newblock \doi{10.1093/mnras/stac2078}.
\newblock URL \url{https://doi.org/10.1093/mnras/stac2078}.

\bibitem[Rezende and Mohamed(2015)]{rezende2015variational}
D.~J. Rezende and S.~Mohamed.
\newblock In \emph{International conference on machine learning}, pages
  1530--1538. PMLR, arXiv, 2015.
\newblock \doi{10.48550/arXiv.1505.05770}.
\newblock URL \url{https://doi.org/10.48550/arxiv.1505.05770}.

\bibitem[Riggi et~al.(2023)Riggi, Magro, Sortino, De~Marco, Bordiu, Cecconello,
  Hopkins, Marvil, Umana, Sciacca, Vitello, Bufano, Ingallinera, Fiameni,
  Spampinato, and Zarb~Adami]{riggiAstronomicalSourceDetection2023}
S.~Riggi et al.
\newblock \emph{Astronomy and Computing}, 42:\penalty0 100682, Jan. 2023.
\newblock ISSN 2213-1337.
\newblock \doi{10.1016/j.ascom.2022.100682}.
\newblock URL \url{https://doi.org/10.1016/j.ascom.2022.100682}.

\bibitem[Riggi et~al.(2024{\natexlab{a}})Riggi, Cecconello, Becciani, and
  Vitello]{riggi_2024_detection_classification}
S.~Riggi, T.~Cecconello, U.~Becciani, and F.~Vitello.
\newblock \emph{CoRR}, 2024{\natexlab{a}}.
\newblock URL \url{http://arxiv.org/abs/2411.08519v1}.

\bibitem[Riggi et~al.(2024{\natexlab{b}})Riggi, Cecconello, Palazzo, Hopkins,
  Gupta, Bordiu, Ingallinera, Buemi, Bufano, Cavallaro, and
  et~al.]{riggi_2024_self_supervised}
S.~Riggi et al.
\newblock \emph{Publications of the Astronomical Society of Australia},
  41:\penalty0 e085, 2024{\natexlab{b}}.
\newblock \doi{10.1017/pasa.2024.84}.
\newblock URL \url{https://doi.org/10.1017/pasa.2024.84}.

\bibitem[Riggi et~al.(2025)Riggi, Cecconello, Pilzer, Palazzo, Gupta, Hopkins,
  Trigilio, and Umana]{riggi_2025_llava}
S.~Riggi et al.
\newblock \emph{Publications of the Astronomical Society of Australia},
  42:\penalty0 e121, 2025.
\newblock ISSN 1448-6083.
\newblock \doi{10.1017/pasa.2025.10082}.
\newblock URL \url{http://dx.doi.org/10.1017/pasa.2025.10082}.

\bibitem[Robert and Casella(2011)]{Robert2011}
C.~Robert and G.~Casella.
\newblock \emph{Statistical Science}, 26\penalty0 (1), Feb. 2011.
\newblock ISSN 0883-4237.
\newblock \doi{10.1214/10-sts351}.
\newblock URL \url{http://dx.doi.org/10.1214/10-STS351}.

\bibitem[Robishaw et~al.(2026)Robishaw, author2, author3, author4, and
  author5]{Robishaw01.2026.SKA}
T.~Robishaw et al.
\newblock In \emph{Advancing Astrophysics with the SKA -- II (AASKAII)}. 2026.
\newblock arXiv search: Report number AASKAII/Robishaw01.

\bibitem[Ronneberger et~al.(2015)Ronneberger, Fischer, and Brox]{unet_2015}
O.~Ronneberger, P.~Fischer, and T.~Brox.
\newblock In N.~Navab, J.~Hornegger, W.~M. Wells, and A.~F. Frangi, editors,
  \emph{Medical Image Computing and Computer-Assisted Intervention -- MICCAI
  2015}, pages 234--241, Cham, 2015. Springer International Publishing.
\newblock ISBN 9783319245744.
\newblock \doi{10.1007/978-3-319-24574-4_28}.
\newblock URL \url{https://doi.org/10.1007/978-3-319-24574-4_28}.

\bibitem[R{\"u}stig et~al.(2024)R{\"u}stig, Guardiani, Roth, Frank, and
  En{\ss}lin]{rustigIntroducingLensCharmCharming2024}
J.~R{\"u}stig et al.
\newblock \emph{A\&A}, 682:\penalty0 A146, Feb. 2024.
\newblock ISSN 0004-6361, 1432-0746.
\newblock \doi{10.1051/0004-6361/202348256}.
\newblock URL \url{https://doi.org/10.1051/0004-6361/202348256}.

\bibitem[Sadr et~al.(2020)Sadr, Bassett, Oozeer, Fantaye, and
  Finlay]{sadr_2020_deep_learning}
A.~V. Sadr et al.
\newblock \emph{Monthly Notices of the Royal Astronomical Society},
  499\penalty0 (1):\penalty0 379--390, 2020.
\newblock \doi{10.1093/mnras/staa2724}.
\newblock URL \url{http://dx.doi.org/10.1093/mnras/staa2724}.

\bibitem[Sasmal et~al.(2026)Sasmal, author2, author3, author4, and
  author5]{Sasmal01.2026.SKA}
T.~Sasmal et al.
\newblock In \emph{Advancing Astrophysics with the SKA -- II (AASKAII)}. 2026.
\newblock arXiv search: Report number AASKAII/Sasmal01.

\bibitem[Scaife(2020)]{scaife_2020_big_telescope}
A.~M.~M. Scaife.
\newblock \emph{Philosophical Transactions of the Royal Society A:
  Mathematical, Physical and Engineering Sciences}, 378\penalty0
  (2166):\penalty0 20190060, 2020.
\newblock \doi{10.1098/rsta.2019.0060}.
\newblock URL \url{http://dx.doi.org/10.1098/rsta.2019.0060}.

\bibitem[Schaeffer et~al.(2023)Schaeffer, Giri, and
  Schneider]{schaeffer_2023_beorn}
T.~Schaeffer, S.~K. Giri, and A.~Schneider.
\newblock \emph{Monthly Notices of the Royal Astronomical Society},
  526\penalty0 (2):\penalty0 2942--2959, 2023.
\newblock \doi{10.1093/mnras/stad2937}.
\newblock URL \url{http://dx.doi.org/10.1093/mnras/stad2937}.

\bibitem[Schmidt et~al.(2022)Schmidt, Geyer, Fr{\"o}se, Blomenkamp,
  Br{\"u}ggen, de~Gasperin, Els{\"a}sser, and
  Rhode]{schmidt_2022_deep_learning}
K.~Schmidt et al.
\newblock \emph{Astronomy \& Astrophysics}, 664:\penalty0 A134, 2022.
\newblock \doi{10.1051/0004-6361/202142113}.
\newblock URL \url{http://dx.doi.org/10.1051/0004-6361/202142113}.

\bibitem[Schosser et~al.(2025)Schosser, Heneka, and
  Plehn]{schosser_2025_optimal_fast}
B.~Schosser, C.~Heneka, and T.~Plehn.
\newblock \emph{SciPost Phys. Core}, 8:\penalty0 037, 2025.
\newblock \doi{10.21468/SciPostPhysCore.8.2.037}.
\newblock URL \url{https://scipost.org/10.21468/SciPostPhysCore.8.2.037}.

\bibitem[Sculley et~al.(2015)Sculley, Holt, Golovin, Davydov, Phillips, Ebner,
  Chaudhary, Young, Crespo, and Dennison]{sculley2015}
D.~Sculley et al.
\newblock In \emph{Proceedings of the 29th International Conference on Neural
  Information Processing Systems - Volume 2}, NIPS'15, page 2503–2511,
  Cambridge, MA, USA, 2015. MIT Press.
\newblock URL
  \url{https://proceedings.neurips.cc/paper/2015/file/86df7dcfd896fcaf2674f757a2463eba-Paper.pdf}.

\bibitem[{Semelin} et~al.(2025){Semelin}, {M\'eriot}, {Mishra}, and
  {Cornu}]{semelin_2025_combining_summary}
B.~{Semelin}, R.~{M\'eriot}, A.~{Mishra}, and D.~{Cornu}.
\newblock \emph{A\&A}, 698:\penalty0 A35, 2025.
\newblock \doi{10.1051/0004-6361/202453115}.
\newblock URL \url{https://doi.org/10.1051/0004-6361/202453115}.

\bibitem[Shannon et~al.(2026)Shannon, author2, author3, author4, and
  author5]{Shannon01.2026.SKA}
R.~M. Shannon et al.
\newblock In \emph{Advancing Astrophysics with the SKA -- II (AASKAII)}. 2026.
\newblock arXiv search: Report number AASKAII/Shannon01.

\bibitem[Sharma et~al.(2026)Sharma, Felix, Valle, Timmel, Gehrig, Wassmer,
  Studer, Hitz, Schramka, Bianco, Crichton, Spinelli, Csillaghy, K{\"o}gel, and
  R{\'e}fr{\'e}gier]{sharma_2026_karabo}
R.~Sharma et al.
\newblock \emph{Astronomy and Computing}, 54:\penalty0 101004, 2026.
\newblock \doi{10.1016/j.ascom.2025.101004}.
\newblock URL \url{http://dx.doi.org/10.1016/j.ascom.2025.101004}.

\bibitem[{SKA Observatory}(2025)]{srcnet_docs_2025}
{SKA Observatory}.
\newblock {SRCNet} documents.
\newblock
  \url{https://confluence.skatelescope.org/display/SNC/SRCNet+Documents}, 2025.

\bibitem[Slijepcevic et~al.(2023)Slijepcevic, Scaife, Walmsley, Bowles, Wong,
  Shabala, and White]{slijepcevic_2023_radio_galaxy_zoo}
I.~V. Slijepcevic et al.
\newblock 2023.
\newblock \doi{10.48550/ARXIV.2305.16127}.
\newblock URL \url{https://arxiv.org/abs/2305.16127}.

\bibitem[Sohl-Dickstein et~al.(2015)Sohl-Dickstein, Weiss, Maheswaranathan, and
  Ganguli]{sohl2015deep}
J.~Sohl-Dickstein, E.~A. Weiss, N.~Maheswaranathan, and S.~Ganguli.
\newblock In \emph{International conference on machine learning}, pages
  2256--2265. PMLR, arXiv, 2015.
\newblock \doi{10.48550/arXiv.1503.03585}.
\newblock URL \url{https://doi.org/10.48550/arXiv.1503.03585}.

\bibitem[Song and Ermon(2019)]{song2019generative}
Y.~Song and S.~Ermon.
\newblock \emph{Advances in neural information processing systems}, 32, 2019.
\newblock \doi{10.48550/arXiv.1907.05600}.
\newblock URL \url{https://doi.org/10.48550/arXiv.1907.05600}.

\bibitem[Song and Ermon(2020)]{song2020improved}
Y.~Song and S.~Ermon.
\newblock \emph{Advances in neural information processing systems},
  33:\penalty0 12438--12448, 2020.
\newblock \doi{10.48550/arXiv.2006.09011}.
\newblock URL \url{https://doi.org/10.48550/arxiv.2006.09011}.

\bibitem[Song and Kingma(2021)]{song2021train}
Y.~Song and D.~P. Kingma.
\newblock \emph{arXiv preprint arXiv:2101.03288}, 2021.
\newblock \doi{10.48550/arXiv.2101.03288}.
\newblock URL \url{https://doi.org/10.48550/arXiv.2101.03288}.

\bibitem[Song et~al.(2020)Song, Sohl-Dickstein, Kingma, Kumar, Ermon, and
  Poole]{song2020score}
Y.~Song et al.
\newblock \emph{arXiv preprint arXiv:2011.13456}, 2020.
\newblock \doi{10.48550/arXiv.2011.13456}.
\newblock URL \url{https://doi.org/10.48550/arxiv.2011.13456}.

\bibitem[Song et~al.(2021)Song, Durkan, Murray, and Ermon]{song2021maximum}
Y.~Song, C.~Durkan, I.~Murray, and S.~Ermon.
\newblock \emph{Advances in Neural Information Processing Systems},
  34:\penalty0 1415--1428, 2021.
\newblock URL \url{https://openreview.net/forum?id=AklttWFnxS9}.

\bibitem[Sortino et~al.(2023)Sortino, Magro, Fiameni, Sciacca, Riggi, DeMarco,
  Spampinato, Hopkins, Bufano, Schillirò, Bordiu, and
  Pino]{sortino_2023_radio_astronomical}
R.~Sortino et al.
\newblock \emph{Experimental Astronomy}, 56\penalty0 (1):\penalty0 293--331,
  2023.
\newblock \doi{10.1007/s10686-023-09893-w}.
\newblock URL \url{http://dx.doi.org/10.1007/s10686-023-09893-w}.

\bibitem[Spinelli et~al.(2026)Spinelli, author2, author3, author4, and
  author5]{Spinelli01.2026.SKA}
M.~Spinelli et al.
\newblock In \emph{Advancing Astrophysics with the SKA -- II (AASKAII)}. 2026.
\newblock arXiv search: Report number AASKAII/Spinelli01.

\bibitem[Sravan et~al.(2022)Sravan, Graham, Fremling, and
  Coughlin]{sravan_2022_autonomous_real_time}
N.~Sravan, M.~J. Graham, C.~Fremling, and M.~W. Coughlin.
\newblock 2022.
\newblock URL \url{https://arxiv.org/abs/2112.05897}.

\bibitem[{SRCNet}(2025)]{srcnet_v01_2025}
{SRCNet}.
\newblock {SRCNet} v0.1 implementation plan.
\newblock Technical Report SRC-0000009-01, Feb 2025.
\newblock URL
  \url{https://docs.google.com/document/d/11DnUf52hDejtUr76DCaCss6OP9K0IEzxer-tUxOkGyM/edit?usp=sharing}.

\bibitem[Sun et~al.(2026)Sun, author2, author3, author4, and
  author5]{Sun01.2026.SKA}
X.~Sun et al.
\newblock In \emph{Advancing Astrophysics with the SKA -- II (AASKAII)}. 2026.
\newblock arXiv search: Report number AASKAII/Sun01.

\bibitem[Sutton et~al.(1998)Sutton, Barto, et~al.]{sutton1998reinforcement}
R.~S. Sutton, A.~G. Barto, et~al.
\newblock \emph{Reinforcement learning: An introduction}, volume~1.
\newblock MIT press Cambridge, 1998.
\newblock \doi{10.1109/tnn.1998.712192}.
\newblock URL \url{https://doi.org/10.1109/tnn.1998.712192}.

\bibitem[Swierc et~al.(2024)Swierc, Tamargo-Arizmendi, {\'C}iprijanovi{\'c},
  and Nord]{swierc_2024_domain_adaptive}
P.~Swierc, M.~Tamargo-Arizmendi, A.~{\'C}iprijanovi{\'c}, and B.~D. Nord.
\newblock \emph{CoRR}, 2024.
\newblock URL \url{http://arxiv.org/abs/2410.16347v1}.

\bibitem[Tahani et~al.(2026)Tahani, author2, author3, author4, and
  author5]{Tahani01.2026.SKA}
M.~Tahani et al.
\newblock In \emph{Advancing Astrophysics with the SKA -- II (AASKAII)}. 2026.
\newblock arXiv search: Report number AASKAII/Tahani01.

\bibitem[Tajja et~al.(2025)Tajja, Aghabiglou, Tolley, Kneib, Thiran, and
  Wiaux]{tajja_2025_s_r2d2}
A.~Tajja et al.
\newblock \emph{Monthly Notices of the Royal Astronomical Society}, page
  staf1082, 2025.
\newblock \doi{10.1093/mnras/staf1082}.
\newblock URL \url{http://dx.doi.org/10.1093/mnras/staf1082}.

\bibitem[Takahashi(2023)]{takahashi_2023_introduction_to}
K.~Takahashi.
\newblock \emph{Publications of the Astronomical Society of Japan},
  75:\penalty0 S50--S84, 2023.
\newblock \doi{10.1093/pasj/psac111}.
\newblock URL \url{http://dx.doi.org/10.1093/pasj/psac111}.

\bibitem[Terris et~al.(2022)Terris, Dabbech, Tang, and
  Wiaux]{terris_2022_image_reconstruction}
M.~Terris, A.~Dabbech, C.~Tang, and Y.~Wiaux.
\newblock \emph{Monthly Notices of the Royal Astronomical Society},
  518\penalty0 (1):\penalty0 604--622, 2022.
\newblock \doi{10.1093/mnras/stac2672}.
\newblock URL \url{http://dx.doi.org/10.1093/mnras/stac2672}.

\bibitem[Tiwari et~al.(2022)Tiwari, Shaw, Majumdar, Kamran, and
  Choudhury]{tiwariImprovingConstraintsReionization2022}
H.~Tiwari et al.
\newblock \emph{J. Cosmol. Astropart. Phys.}, 2022\penalty0 (04):\penalty0 045,
  Apr. 2022.
\newblock ISSN 1475-7516.
\newblock \doi{10.1088/1475-7516/2022/04/045}.
\newblock URL \url{https://doi.org/10.1088/1475-7516/2022/04/045}.

\bibitem[Toennies(2024)]{toennies2024introduction}
K.~D. Toennies.
\newblock \emph{An Introduction to Image Classification: From Designed Models
  to End-to-End Learning}.
\newblock Springer Nature Singapore, 2024.
\newblock ISBN 9789819978823.
\newblock \doi{10.1007/978-981-99-7882-3}.
\newblock URL \url{https://doi.org/10.1007/978-981-99-7882-3}.

\bibitem[Tolley(2024)]{tolleyWaveletScatteringNetworks2024}
E.~Tolley.
\newblock \emph{RSL}, 5, 2024.
\newblock \doi{10.46620/23-0028}.
\newblock URL \url{https://doi.org/10.46620/23-0028}.

\bibitem[Tolley et~al.(2025)Tolley, Frasch, Orliac, Krishna, Bianco, Kashani,
  Hurley, Simeoni, and Kneib]{tolley_2025_bipp}
E.~Tolley et al.
\newblock \emph{Astronomy and Computing}, 51:\penalty0 100920, 2025.
\newblock \doi{10.1016/j.ascom.2024.100920}.
\newblock URL \url{http://dx.doi.org/10.1016/j.ascom.2024.100920}.

\bibitem[Tripathi et~al.(2025)Tripathi, Datta, Mazumder, and
  Majumdar]{tripathiImpactCalibrationPosition2025a}
A.~Tripathi, A.~Datta, A.~Mazumder, and S.~Majumdar.
\newblock \emph{J. Cosmol. Astropart. Phys.}, 2025\penalty0 (10):\penalty0 035,
  Oct. 2025.
\newblock ISSN 1475-7516.
\newblock \doi{10.1088/1475-7516/2025/10/035}.
\newblock URL \url{https://doi.org/10.1088/1475-7516/2025/10/035}.

\bibitem[Trott et~al.(2026)Trott, author2, author3, author4, and
  author5]{Trott01.2026.SKA}
C.~Trott et al.
\newblock In \emph{Advancing Astrophysics with the SKA -- II (AASKAII)}. 2026.
\newblock arXiv search: Report number AASKAII/Trott01.

\bibitem[Van Den~Oord et~al.(2016)Van Den~Oord, Kalchbrenner, and
  Kavukcuoglu]{van2016pixel}
A.~Van Den~Oord, N.~Kalchbrenner, and K.~Kavukcuoglu.
\newblock In \emph{International conference on machine learning}, pages
  1747--1756. PMLR, arXiv, 2016.
\newblock \doi{10.48550/arXiv.1601.06759}.
\newblock URL \url{https://doi.org/10.48550/arxiv.1601.06759}.

\bibitem[Vaswani et~al.(2017)Vaswani, Shazeer, Parmar, Uszkoreit, Jones, Gomez,
  Kaiser, and Polosukhin]{vaswani2017}
A.~Vaswani et al.
\newblock In I.~Guyon et al., editors, \emph{Advances in Neural Information
  Processing Systems}, volume~30. Curran Associates, Inc., 2017.
\newblock URL
  \url{https://proceedings.neurips.cc/paper_files/paper/2017/file/3f5ee243547dee91fbd053c1c4a845aa-Paper.pdf}.

\bibitem[Vernstrom et~al.(2026)Vernstrom, West, and
  Horellou]{Vernstrom01.2026.SKA}
T.~Vernstrom, J.~L. West, and C.~Horellou.
\newblock In \emph{Advancing Astrophysics with the SKA -- II (AASKAII)}. 2026.
\newblock arXiv search: Report number AASKAII/Vernstrom01.

\bibitem[Wang et~al.(2026)Wang, author2, author3, author4, and
  author5]{JingWang01.2026.SKA}
J.~Wang et al.
\newblock In \emph{Advancing Astrophysics with the SKA -- II (AASKAII)}. 2026.
\newblock arXiv search: Report number AASKAII/JingWang01.

\bibitem[Wang et~al.(2023)Wang, Chen, Zhu, Luo, and Wang]{wang_2023_polarrec}
R.~Wang et al.
\newblock \emph{CoRR}, 2023.
\newblock URL \url{http://arxiv.org/abs/2308.14610v2}.

\bibitem[Wang et~al.(2025{\natexlab{a}})Wang, Wang, Luo, Wang, and
  Wu]{wang_2025_visrec}
R.~Wang et al.
\newblock \emph{Proceedings of the AAAI Conference on Artificial Intelligence},
  39\penalty0 (1):\penalty0 852--860, 2025{\natexlab{a}}.
\newblock \doi{10.1609/aaai.v39i1.32069}.
\newblock URL \url{http://dx.doi.org/10.1609/aaai.v39i1.32069}.

\bibitem[Wang et~al.(2025{\natexlab{b}})Wang, Mignot, Prunet, Mascolo,
  Spinelli, and Ferrari]{wang_2025_decentralized_framework}
S.~Wang et al.
\newblock \emph{The Astronomical Journal}, 169\penalty0 (6):\penalty0 289,
  2025{\natexlab{b}}.
\newblock \doi{10.3847/1538-3881/adc57b}.
\newblock URL \url{http://dx.doi.org/10.3847/1538-3881/adc57b}.

\bibitem[Wright et~al.(2024)Wright, Ad{\'a}mek, and
  Armour]{wright_2024_clean_algorithm}
D.~Wright, K.~Ad{\'a}mek, and W.~Armour.
\newblock \emph{CoRR}, 2024.
\newblock URL \url{http://arxiv.org/abs/2403.03091v1}.

\bibitem[Xiao et~al.(2021)Xiao, Kreis, and Vahdat]{xiao2021tackling}
Z.~Xiao, K.~Kreis, and A.~Vahdat.
\newblock \emph{arXiv preprint arXiv:2112.07804}, 2021.
\newblock \doi{10.48550/arXiv.2112.07804}.
\newblock URL \url{https://doi.org/10.48550/arXiv.2112.07804}.

\bibitem[Xu et~al.(2025)Xu, Karcheski, Law, Zhu, Hsu, and
  Tan]{xu_2025_exploring_magnetic}
D.~Xu et al.
\newblock \emph{The Astrophysical Journal}, 980\penalty0 (1):\penalty0 52,
  2025.
\newblock \doi{10.3847/1538-4357/ada8a0}.
\newblock URL \url{http://dx.doi.org/10.3847/1538-4357/ada8a0}.

\bibitem[Yang et~al.(2023)Yang, Zhang, Song, Hong, Xu, Zhao, Zhang, Cui, and
  Yang]{yang2023diffusion}
L.~Yang et al.
\newblock \emph{ACM Computing Surveys}, 56\penalty0 (4):\penalty0 1--39, Nov.
  2023.
\newblock ISSN 1557-7341.
\newblock \doi{10.1145/3626235}.
\newblock URL \url{https://doi.org/10.1145/3626235}.

\bibitem[Yatawatta(2023)]{yatawatta_2023_hint_assisted}
S.~Yatawatta.
\newblock \emph{CoRR}, 2023.
\newblock URL \url{http://arxiv.org/abs/2301.03933v1}.

\bibitem[Zhang et~al.(2026)Zhang, author2, author3, author4, and
  author5]{LeZhang01.2026.SKA}
L.~Zhang et al.
\newblock In \emph{Advancing Astrophysics with the SKA -- II (AASKAII)}. 2026.
\newblock arXiv search: Report number AASKAII/LeZhang01.

\bibitem[Zhang et~al.(2024)Zhang, Cognard, and Dobigeon]{zhang_2024_rfi_drunet}
X.~Zhang, I.~Cognard, and N.~Dobigeon.
\newblock \emph{Astronomy and Computing}, 47:\penalty0 100822, 2024.
\newblock ISSN 2213-1337.
\newblock \doi{https://doi.org/10.1016/j.ascom.2024.100822}.
\newblock URL
  \url{https://www.sciencedirect.com/science/article/pii/S2213133724000374}.

\bibitem[Zhang et~al.(2022)Zhang, Jiang, and
  Zhang]{zhang_2022_automatic_detection}
Z.~Zhang, B.~Jiang, and Y.~Zhang.
\newblock \emph{Publications of the Astronomical Society of the Pacific},
  134\penalty0 (1036):\penalty0 064503, 2022.
\newblock \doi{10.1088/1538-3873/ac67b1}.
\newblock URL \url{http://dx.doi.org/10.1088/1538-3873/ac67b1}.

\bibitem[Zhu et~al.(2017)Zhu, Park, Isola, and Efros]{zhu2017unpaired}
J.-Y. Zhu, T.~Park, P.~Isola, and A.~A. Efros.
\newblock In \emph{Proceedings of the IEEE international conference on computer
  vision}, pages 2223--2232. IEEE, Oct. 2017.
\newblock \doi{10.1109/iccv.2017.244}.
\newblock URL \url{https://doi.org/10.1109/iccv.2017.244}.

\end{thebibliography}

\end{document}